\documentclass[journal]{IEEEtran}
  
\usepackage{biblatex}
\usepackage{csquotes}
\usepackage{amsmath,amssymb,amsfonts}
\usepackage{algorithmic}
\usepackage{graphicx}
\usepackage{textcomp}
\usepackage{tabularx}
\usepackage{rotating}
\usepackage{adjustbox}
\usepackage{url}
\usepackage{pdflscape}
\usepackage{multirow, tabularx}
\usepackage[super]{nth}
\usepackage{xcolor}
\def\BibTeX{{\rm B\kern-.05em{\sc i\kern-.025em b}\kern-.08em
   T\kern-.1667em\lower.7ex\hbox{E}\kern-.125emX}}

\usepackage{amsmath}
\DeclareUnicodeCharacter{0301}{\'{e}}
\usepackage{enumitem}
\usepackage{makecell}
\usepackage{hyperref}
\hypersetup{
    colorlinks=false,
    linkcolor=blue,
    filecolor=magenta,      
    urlcolor=blue,
    pdftitle={Sharelatex Example},
    bookmarks=false,
    pdfpagemode=FullScreen,
    }
\usepackage{csquotes}

\begin{document}

\title{A Survey of Privacy Vulnerabilities
\\of Mobile Device Sensors
}

\author{\IEEEauthorblockN{Paula Delgado-Santos}\IEEEauthorrefmark{1}\IEEEauthorrefmark{2}, 
Giuseppe Stragapede\IEEEauthorrefmark{1}\IEEEauthorrefmark{3},
Ruben Tolosana\IEEEauthorrefmark{3},\\
Richard Guest\IEEEauthorrefmark{2},
Farzin Deravi\IEEEauthorrefmark{2},
Ruben Vera-Rodriguez\IEEEauthorrefmark{3}\\
\IEEEauthorblockA{\IEEEauthorrefmark{2}School of Engineering and Digital Arts, University of Kent \\ Email: (p.delgado-de-santos, r.m.guest, f.deravi@kent.ac.uk)@kent.ac.uk} \\
\IEEEauthorblockA{\IEEEauthorrefmark{3}Biometrics and Data Pattern Analytics Lab, Universidad Autonoma de Madrid \\ Email: (giuseppe.stragapede, ruben.tolosana, ruben.vera)@uam.es}}



\maketitle

\begin{abstract}

The number of mobile devices, such as smartphones and smartwatches, is relentlessly increasing to almost 6.8 billion by 2022, and along with it, the amount of personal and sensitive data captured by them. This survey overviews the state of the art of what personal and sensitive user attributes can be extracted from mobile device sensors, emphasising critical aspects such as demographics, health and body features, activity and behaviour recognition, etc. In addition, we review popular metrics in the literature to quantify the degree of privacy, and discuss powerful privacy methods to protect the sensitive data while preserving data utility for analysis. Finally, open research questions are presented for further advancements in the field.
\end{abstract}


\def\thefootnote{*}\footnotetext{These authors contributed equally to this paper.}
\def\thefootnote{\arabic{footnote}}

\section{Introduction}
\label{sec:introduction}
Mobile devices such as smartphones, tablets, and wearables are provided with several sensors that are able to acquire a vast amount of personal information in different forms and for different purposes. This aspect, in combination with the significant advancements of the computational and communication capabilities of mobile devices over the last years, has shown the high potential of mobile devices in many application fields~\cite{Abuhamad2020Sensorbased, ellavarason2020touch, tolosana2021child}.
The large availability of personal data generated on mobile devices, in combination with their ubiquity (with 3.9 billion smartphones globally in 2016, estimated to rise to 6.8 billion by 2022~\cite{Heuveldop2017}) and their \textit{always-on} nature has turned this technology into a potential source of major invasion of persona privacy. 
\par The European Union has provided the General Data Protection Regulation (GDPR), defining personal data as any information related to an identified or identifiable natural person~\cite{GDPR}. Moreover, the GDPR also defines sensitive data as a subset of personal information, that includes: \textit{i)} personal data revealing racial or ethnic origin, political opinions, religious or philosophical beliefs; \textit{ii)} trade-union membership; \textit{iii)} genetic data, biometric data processed solely to identify a human being; \textit{iv)} health-related data; and \textit{v)} data concerning a person’s sex life or sexual orientation \cite{GDPR}. Automated processing of user data, also known as user profiling \cite{GDPR}, can easily reveal such attributes from data acquired through mobile user interaction by requesting irrelevant permissions, lax definition of permissions or misuse of permissions, combined with the aggregation of highly personalised data, reducing the privacy and security experience of the final users~\cite{Barth2019}. Consequently, many works in the literature have focused on preventing potential misuse. This is the motivation of recent EU-funded, Innovative Training Networks (ITN) such as PriMa\cite{PriMa} and TReSPAsS \cite{TReSPAsS}. 
\par In this context, we distinguish between privacy protection and sensitive data protection. They both aim to de-identify the user data and avoid re-identification \cite{garfinkel2015identification} of direct identifiers, such as names, social security numbers, addresses, etc. \cite{ISO25237}, and indirect identifiers. The latter are not capable of identifying a particular individual but can be used in conjunction with other information to identify data subjects \cite{dalenius1986finding}. However, privacy protection refers to the security of the personal data and it borrows terminology, definitions and methods from cybersecurity, whereas sensitive data protection focuses on data modification techniques that account for the sensitive data while maximising the residual data utility for analysis. The idea of selective sensitive data protection was conceived with the development of the first large databases \cite{10.1145/170035.170072}. Early works in this field led to the concept of data sanitisation \cite{836532}, as a database transformation before its release to a third party, and to the concept of Privacy Preserving Data Mining (PPDM) \cite{agrawal2000privacy}, as the development of models about aggregated data without access to precise information in individual data records. Furthermore, the term de-identification was coined to define the operation of Personal Identifiable Information\footnote{Personal Identifiable Information is defined as information sufficient to distinguish or trace an individual's identity. This information may be used on its own or in conjunction with other information relating to an individual \cite{krishnamurthy2009}.} (PII) removal from data collected, used, archived, and shared by organisations \cite{garfinkel2015identification}.



\par Privacy is a multifaceted concept which has received a plethora of formulations and definitions~\cite{RutgersLawReview, de2012data, Dwork2014, Barker2009}. A profound discussion of the concept of privacy is, however, not in the scope of the present work. We will adopt the perspective of Article 21 of the GDPR, which states that the subject shall have the right to object, on grounds relating to his or her particular situation, at any time to processing of personal data concerning him or her. From this perspective, the main contributions of the present article are:

\begin{itemize}
\item An overview of the sensors and the raw data commonly available in modern mobile devices, paying special attention to the background sensors as they may be considered innocuous by the end users.

\item A description of the typical application scenarios and purposes of collected data for mobile scenarios.

\item An in-depth analysis of the personal and sensitive data extracted from mobile background sensors and the corresponding automated methods, focusing on: demographics, activity and behaviour, health parameters and body features, mood and emotion, location tracking, and keystroke logging.

\item A summary of the metrics proposed in the literature for privacy quantification from the perspective of sensitive data, including also a review of the methods to achieve sensitive data protection. 
\end{itemize}



\par For completeness, we would like to highlight other recent surveys in the field focusing on other privacy aspects. In~\cite{HernandezAlvarez2021}, the authors focused on privacy protection in the context of authentication. In \cite{Haris2014}, a broad survey was presented about privacy leakage in mobile computing with especial interest in mobile applications, advertising libraries, and connectivity. A comprehensive overview is provided, but without a specific focus on the sensitive data. Finally, an analysis of privacy in the context of soft biometrics\footnote{\textit{Soft biometrics} is defined as the characteristics that provide some information about the individual, but lack the distinctiveness and permanence to sufficiently differentiate any two individuals~\cite{Jain2004a}.} is considered in \cite{Dantcheva2016}, focusing on the extraction of demographic information such as gender from image and video data, not from mobile background sensors as in the present survey. Similarly, privacy was studied in the context of audio data in \cite{Gao2019}. In contrast to previous work, we pay special attention to sensitive data and provide, to the best of our knowledge, the first survey that focuses directly on sensitive data and their privacy protection metrics and methods.


\par The rest of this survey is organised as follows. We first provide in Sec. \ref{sec:mobile_acquisition_sensitive_data} an overview of the sensors and the raw data commonly available in modern mobile devices. In Sec. \ref{sec:sensor_application_scenarios} the typical application scenarios and purposes of collected data are described. Sensitive user information extraction is addressed in Sec. \ref{sec:privacy_sensitive_data}. In addition, the methods used to achieve the goal of extracting information are systematically discussed. Sec. \ref{sec:privacy_metrics} focuses on metrics whereas Sec. \ref{sec:privacy_protection_methods} focuses on methods for privacy protection techniques from the perspective of the sensitive data. In Sec. \ref{sec:conclusion_and_open_directions}, the general conclusions of the present study are drawn and some open research questions that have emerged through the present survey are outlined for further investigation.

 
\section{Mobile Acquisition of Sensitive Data}
\label{sec:mobile_acquisition_sensitive_data}
Mobile devices offer a rich ground for data collection and processing. Smartphones, to begin with, are in fact distinguished from previous generation cellular phones by their stronger hardware capabilities (e.g., equipped with multi-core processors, GPUs, hardware acceleration units and gigabytes of memory) and powerful mobile operating systems, which facilitate wider sensing, software, internet, and multimedia functionalities, alongside core phone functions. 
\par Mobile device built-in sensors, known as background sensors, are capable of providing frequent measures of physical quantities in an unobtrusive and transparent way. However, these data can be easily utilised to extract sensitive information of the user such as gender, age, emotion, ethnic group, etc. 
\par This is also the case of other popular wearable devices such as smartwatches. Wearables might be considered under the broad definition of Internet of Things (IoT) devices since they are connected to the internet to collect and exchange data to perform automated decision making \cite{9202574}. Their popularity among consumer electronics is rapidly increasing and they are progressively becoming capable of more specialised measurements and analyses \cite{Dian2020}. In general, wearable manufacturers often provide users with mobile applications to install on their smartphones for communication and computing purposes, together with a more complete user interface. For example, smartwatches or fitness tracker bracelets can provide measurements of walked or run distances (based on data from motion sensors and Global Positioning System (GPS)) but also physiological parameters such as heart rate, Electrocardiogram (ECG), stress, sleep quality, etc. 
\par Table \ref{data_acquisition_table} provides a description of the sensors and the raw data commonly available in modern mobile devices, grouped according to their sensing domain. In general, sensors can be classified into two categories based on the process adopted to produce the output signal: \textit{i)} hardware sensors, on one hand, are physically installed components that perform a transduction of the physical quantity they measure to an electrical signal, which is converted into the digital domain for further processing; \textit{ii)} software sensors, on the other hand, rely on data already made available by hardware sensor and/or calculate them to produce a measurement. 

\begin{table*}[t]\caption{Description of the sensors and the raw data commonly available in modern mobile devices. BPM- Beats Per Minute, ECG- Electrocardiogram, SpO$_2$- Saturation of Peripheral Oxygen, GPS- Global Positioning System, SSID- Service Set IDentifier, RSSI- Receiver Signal Strength Indicator.} 
\centering
\label{data_acquisition_table}
\begin{tabular}{| l | l l l l |}
\hline
\textbf{Sensor Type} & \makecell[l]{\textbf{Sensor /}\\ \textbf{Data Source}} & \makecell[l]{\textbf{Measured /}\\ \textbf{Logged Quantity}} & \textbf{Scope / Purpose} & \makecell{\textbf{Sensor Type}}\\
\hline
\hline
\multicolumn{1}{|l|}{\multirow{10}{*}{Motion}} & \makecell[l]{Accelerometer /\\ Linear Accelerometer} & Acceleration Force & Device Traslation & Hardware \\
& Gyroscope & Angular Velocity & Device Rotation & Hardware \\
& \makecell[l]{Rotation Vector} & Angle & Device Orientation & \makecell[l]{Hardware,\\ Software} \\
& Gravity & Magnitude of Gravity & Device Orientation & \makecell[l]{Hardware,\\ Software} \\
& Significant Motion & \makecell[l]{Change of user movement} & \makecell[l]{Walking or Riding Vehicle} & Software \\
& \makecell[l]{Step Counter} & Number of Steps  & Physical Activity Tracking & Software \\
& \makecell[l]{Step Detector} & Step & Physical Activity Tracking & Software \\\hline
\multicolumn{1}{|l|}{\multirow{7}{*}{Position}} & Geomagnetic Field & Earth's Magnetic Field  & Device Orientation & Hardware \\
& Proximity & Distance & Device Distance from Surface & Hardware \\
& Magnetometer & Earth's Magnetic Field & Device Orientation & Hardware\\
& \makecell[l]{Geomagnetic Rotation Vector} & Earth's Magnetic Field & Device Orientation &  \makecell[l]{Hardware,\\ Software}  \\
& \makecell[l]{Game Rotation Vector} & Angle & Device Rotation & \makecell[l]{Hardware,\\ Software}\\\hline
\multicolumn{1}{|l|}{\multirow{4}{*}{Environmental}} & Light & Illuminance  & Screen Luminosity Regulation & Hardware \\
& Pressure & Ambient Pressure & Contextual Information &  Hardware \\
& Temperature & Ambient Temperature & Contextual Information &   Hardware \\
& Humidity & Ambient Humidity  & Contextual information &  Hardware \\
\hline
\multicolumn{1}{|l|}{\multirow{12}{*}{Health}} & BPM & Number of Beats & \makecell[l]{Physical Activity Monitoring} & Hardware \\
& ECG & Sinus Rhythm Graph & \makecell[l]{Physical Activity Monitoring} & Hardware \\
& SpO$_2$ & \makecell[l]{Arterial Blood Oxygen\\ Saturation Percentage Level} & \makecell[l]{Physical Activity Monitoring} & Hardware \\
& Blood Pressure & \makecell[l]{Systolic and Diastolic\\ Average Pressure} & \makecell[l]{Physical Activity Monitoring} & Software \\
& Stress & \makecell[l]{Percentage based on\\ Heart Beat Variability} & \makecell[l]{Physical Activity Monitoring} & Software \\
& Sleep / Wake Amount & Time & \makecell[l]{Physical Activity Monitoring} & \makecell[l]{Hardware,\\  Software} \\
& Sleep Phase Transitions & \makecell[l]{Time} & \makecell[l]{Physical Activity Monitoring}  & \makecell[l]{Hardware,\\  Software} \\
& Caloric Consumption & Step Counter & \makecell[l]{Physical Activity Monitoring} & Software \\
\hline
\multicolumn{1}{|l|}{\multirow{2}{*}{Touchscreen}} & Keystroke & \makecell[l]{Keys Presses and\\ Releases} & \multicolumn{1}{l}{Key Input} & Hardware \\
& Touch Data & \makecell[l]{Screen Coordinates,\\ Pressure of Touch} & \multicolumn{1}{l}{\makecell[l]{Complex Touch Gestures}} & Hardware\\
\hline
\multicolumn{1}{|l|}{\multirow{5}{*}{\makecell[l]{Network, Location\\and Application}}} & Wi-Fi & \makecell[l]{SSID, RSSI, Encryption Protocol,\\ Frequency, Channel} & \multicolumn{1}{l}{Connectivity} &  Hardware \\
& Bluetooth  & \makecell[l]{SSID, RSSI, Encryption Protocol,\\ Frequency, Channel} & \multicolumn{1}{l}{Connectivity} &  Hardware\\
& Cell Tower & ID & \multicolumn{1}{l}{Connectivity} &  Hardware\\
& GPS & \makecell[l]{Latitude, Longitude, Altitude,\\ Bearing, Accuracy} & \multicolumn{1}{l}{Navigation} & Hardware \\
& App Usage & Name and Time of Used Apps & \multicolumn{1}{l}{System Log} & Software \\
\hline


\end{tabular}
\end{table*}

\par Motion sensors are responsible for measuring the acceleration and rotational forces in the three axes of the device. Hardware- based motion sensors will register continuous quantities as in the case of acceleration or angular velocity, whereas, when software-based, their output could be either continuous or event-driven as in the case of a step detector. Position sensors range from measuring changes in the Earth's magnetic field for orientation in space to proximity sensors, whereas environmental sensors are generally triggered by an event and return a single scalar value measurement. When designed to return continuous measurements, the sampling rate of these sensors can reach up to around 200Hz. Nevertheless, their power consumption is low \cite{acien2020becaptcha}. 
\par Specific physiological/biological parameter measurements are also available on many mobile devices thanks to dedicated health sensors. For example, most smartphones and smartwatches include built-in optical sensors to capture changes in blood volume in the arteries under the skin, from which heart-related as well as polysomnographic parameters can be obtained \cite{Tayfur2019, Chen2013}. 
\par Touchscreen data can be in the form of keystrokes acquired from the keyboard \cite{7738412}, or in the form of touch data acquired throughout the user interaction \cite{8998358}. In the former case, the virtual keys pressed are logged together with pressure and timestamps, for each key press and release. From these raw data, it is possible to extract more complex features, such as the hold time, inter-press time, inter-release time, etc. \cite{acien2021typenet}. 
In addition to keystroke, touchscreen panels significantly enlarged the input data space including touch data. In fact, it is possible to track the touch position in terms of \textit{X} and \textit{Y} coordinates in the screen reference system, but also pressure information and complex multi-touch gestures such as swipe, pinch, tap and scroll. Other complex features that can be extracted from touch data are velocity, acceleration, angle and trajectory \cite{tolosana2020exploiting}.
\par Connectivity is yet another fundamental aspect of mobile devices. Their usefulness and ubiquity stem from the vast spectrum of functionalities they support thanks to many installed network protocols. Network connection data retains information about users' routine patterns, therefore it can be used for behavioural profiling and sensitive information extraction \cite{li2018studying}. With the fifth-generation standard for cellular networks (5G) being commercialised and the sixth generation (6G) in development, significant improvement in terms of bit rates and latency will allow for extensive machine-to-machine communications, thus increasing the vast spectrum of functionalities already supported by mobile devices \cite{8412482}.

\section{Sensor Application Scenarios}
\label{sec:sensor_application_scenarios}

In 2008, the two most common mobile operative systems, Android and iOS, had less than 500 apps available for download. To date, Android users are able to download over 2.87 million apps, followed by the Apple App Store with almost 1.96 million apps \cite{appno}. 
The possible application scenarios are wide ranging. 
Here we describe some popular application scenarios using mobile sensors.

\subsection{User Authentication}
\label{subsec:user_authentication}
In traditional authentication schemes, the legitimate user is expected to have knowledge of a secret such as a PIN code, a password or a pattern to gain access (authentication based on ``what-you-know"), or an object, such as a card reader (authentication based on ``what-you-have"), whereas recent authentication schemes largely deployed on mobile devices are based on the ``what-you-are" paradigm: some traits of the user are acquired and processed in order to verify their identity \cite{OGorman2003}. With regard to mobile user authentication, a common approach is based on biometrics (both physiological and behavioural) \cite{JAIN201680}, as in the case of entry-point fingerprint or face-based identification. A severe limitation of these processes consists in the fact that once the device is unlocked, as long as it remains active, an intruder would have unlimited time at their disposal. To provide prolonged protection, several studies have investigated and proved the feasibility of continuous authentication schemes for mobile devices based on behavioural biometrics \cite{Patel2016ContinuousUser}. In this case, biometric data would be continuously acquired in a passive way throughout normal device usage to constantly verify the user's traits. 
Often combined, background sensors \cite{acien2019multilock}, touchscreen \cite{9304858}, and network information \cite{li2018studying} are among the most frequent modalities explored to develop behavioural biometric continuous authentication systems. 

\subsection{Healthcare and Fitness}
\label{subsec:healthcare_and_fitness}
\par Healthcare is a major field of study for mobile applications. The term ``mHealth" was coined to indicate a sub-set of eHealth that includes medical and public health practice supported by mobile devices. Mobile apps help improve healthcare delivery processes and patients could benefit in terms of monitoring and treatment of diseases and chronic conditions, among many other healthcare purposes\cite{Nussbaum2019}. 
Examples of mobile apps include those that provide measurements of postures, report on mental disorders \cite{Gravenhorst2015}, and assess symptoms of conditions such as Parkinson disease, stress, dementia, etc.~\cite{Majumder2019SmartphoneSensors, 
faundez2020handwriting}. Moreover, mobile health apps can be essential in sustaining a healthy lifestyle among people by monitoring and recommending behaviour corrections. From this perspective, mobile devices such as smartwatches are largely used for fitness tracking. Physical exercise monitoring takes place by acquiring and processing background and GPS sensor data in a explicit and transparent way for the user\cite{Antar2019, Anjum2013}. 

\subsection{Location-based Services}
\label{subsec:location_services}
GPS and geolocation data are used by applications to present information related to the environment and the position of the users, for purposes such as targeted advertising, navigation, and recommendations \cite{Haris2014}. These location-aware applications are under the context awareness paradigm \cite{Saha2003}. Additionally, besides their native scope of communication, short range protocols such as Bluetooth and Wi-Fi allow mobile devices to exploit the information of nearby devices for purposes similar to the ones described. This concept can be particularly useful defining a semantic context of immediate surrounding, especially in the case of indoor environments. For example, in \cite{Dierna2016}, the authors explored the feasibility of creating virtual tours in museums or expositions to deliver information about the items in the proximity of the users, who can receive this information on their mobile devices.

\subsection{Other Applications}
\label{subsec:other_applications}
Traditionally, background sensors contribute to improving the mobile device user experience in several ways. For instance, position sensors are useful for recognising the orientation of the device in order to switch from portrait to landscape modality, and vice versa. Light sensor information about the illuminance is used to automatically adjust the screen brightness. The proximity sensor will lock the screen and activate a different speaker when the user is placing a call. Mobile device background sensors are also widely employed for Augmented Reality (AR) applications in several fields, such as education, entertainment, commerce, and navigation, among others \cite{kim2017augmented}. AR-based apps heavily rely on the information provided by the background sensors to deliver information.    
\par In addition, the sophisticated sensing capabilities of mobile devices, combined with their vast diffusion, have led to the idea of accomplishing large-scale sensing through them, known in literature as \textit{mobile participatory sensing} \cite{Burke2006}. Individuals with sensing and computing devices volunteer to collectively share data to measure and map phenomena of common interest, in a crowd-sourced fashion \cite{Haris2014}. Applications where mobile participatory sensing has been used include noise pollution monitoring, litter monitoring, monitoring of traffic and road conditions, among others \cite{melo2017towards}.

\section{Privacy Sensitive Data}
\label{sec:privacy_sensitive_data}
The automated processing of user data acquired by mobile device sensors can reveal a significant amount of personal and sensitive information. In particular, while sensors such as cameras, GPS, or microphone are privacy-sensitive and require explicit user permission, many other sources such as accelerometer, touchscreen or network connection logs are less protected in terms of privacy. However, these data can also become crucial in obtaining private user information, since they can be processed to ascertain attributes that allow to re-identify a person, to extract demographic information or data related to their activity and health, among others. 
\par Processing data from which it is possible to extract personal and sensitive information can lead to problems arising from the nature of these data. A common characteristic of sensitive data is in fact its uniqueness for each individual and its strict association to their owner. 
These implications are particularly relevant with regard to biometric data. 
In the biometric scenario, additional risk factors include: the modalities used to store personal data, the owner of the system, the used recognition modality (authentication or identification in a biometric database), the durability and class of the used traits, depending on which the severity of the consequences can vary \cite{labati2011biometric}.

\begin{table*}[htbp!]
\centering
\caption{Comparison of different state-of-the-art sensitive data acquisition approaches. \textit{k}-NN- \textit{k}-Nearest Neighbours, RF- Random Forest, SVM- Support Vector Machines, LSTM- Long-Short Term Memory, HMM- Hidden Markov Model, AUD- Active User Detection, DBSCAN- Density-based Spatial Clustering of Applications, CNN- Convolutional Neural Network, RNN- Recurrent Neural Network, DTW- Dynamic Time Warping, Acc- Accuracy, ERR-Equal Error Rate, AUROC- Area Under ROC, KL-Score- Kullback-Leibler Score }
\label{tab:TechniquesDatasanitation}
\begin{tabular}{|l|l|l|l|l|}
\hline
\multicolumn{1}{|c|}{\textbf{Sensitive Data}}        & \multicolumn{1}{c|}{\textbf{Sensors}}              & \textbf{Study}                                                                                      & \textbf{Classifier}                    & \textbf{Best Performace}    \\ \hline\hline
\multirow{14}{*}{Demographics}                       & \multirow{7}{*}{Motion}                            & Jain and Kanhangad (2016) \cite{Jain2016InvestigatingGender}             & SVM                                   &  Acc. = 76.83\%              \\ \cline{3-5} 

                                                     &                                                    &       Davarci \textit{et al.} (2017) \cite{Davarci2017AgeGroup}             &   \textit{k}-NN                                  & Acc. =  85.3\%             \\ \cline{3-5}

                                                     &                                                    & Nguyen \textit{et al.} (2019) \cite{nguyen2019kid}                & RF                                     & Acc. = 96\%                \\ \cline{3-5} 

                                                     &                                                    & Singh \textit{et al.} (2019) \cite{Singh2019SideChannel}          & 4 Classifiers                          & Acc. = 80\%                \\ \cline{3-5} 
                                                     &                                                    & Sabir \textit{et al.} (2019) \cite{sabir2019gait}                 & LSTM + Leave One Out                    & Acc. =  94.11\%            \\ \cline{3-5} 

                                                     &                                                    & Ngo \textit{et al.} (2019) \cite{Ngo2019OUISIR}                   & HMM                                    & ERR = 5.39\%               \\ \cline{3-5} 
                                                     &                                                    & Meena and Saeawadekar (2020) \cite{meena2020gender}                                & Ensemble Boosted Tree                  & Acc. = 96.3\%              \\ \cline{2-5} 
                                                    &  \multirow{4}{*}{Touchscreen}               & Miguel-Hurtado \textit{et al.} (2016) \cite{miguel2016predicting} & Decision Tree & Acc. = 78\%                \\ \cline{3-5} 
                                                     &                       & Acien \textit{et al.} (2019) \cite{Acien2019ActiveDetection}      & AUD                                    & Acc. =  97\%               \\ \cline{3-5} 

                                                     &                                                    & Nguyen \textit{et al.} (2019) \cite{nguyen2019kid}                & RF                                     & Acc. = 99\%                \\ \cline{3-5}

                                                     &                                                    & Jain and Kanhangad (2019) \cite{jain2019gender}                                    & \textit{k}-NN                                   & Acc. = 93.65\%             \\ \cline{2-5} 
                                                    
                                                     &  \multirow{3}{*}{\makecell[l]{Network, Location\\and Application}}               & Riederer \textit{et al.} (2015) \cite{Riederer2015}               & Logistic Regresion                     & Acc. = 72\%                \\ \cline{3-5} 
                                                     &                                                    & Neal and Woodard (2018) \cite{neal2018gender}                                      & RF + Na\"{i}ve Bayes                       & Acc. = 91.8\%   
                                                     \\ \cline{3-5}
                                                      &  & Wu \textit{et al.} (2019) \cite{Wu2019}                           & XGBoost                                & Acc. = 80\%                 \\ \hline
\multirow{8}{*}{Activity and Behaviour}              & \multirow{5}{*}{Motion}                            & Sun \textit{et al.} (2010) \cite{Sun2010Activity}                 & SVM                                    & Acc. = 93.2\%              \\ \cline{3-5} 
                                                     &                                                & Anjum and Ilyas (2013) \cite{anjum2013activity}                                    & Decision Tree                          & AUROC = 99\%               \\ \cline{3-5} 
                                                     &                                                    & Thomaz \textit{et al.} (2015) \cite{thomaz2015practical}          & DBSCAN                                 & Acc. =  76.1\%              \\ \cline{3-5} 
                                                     &                                                    & Arnold \textit{et al.} (2015) \cite{Arnold2015}                   & RF                                     & Acc. = 70\%                \\ \cline{3-5} 
                                                     &                                                    & Chang \textit{et al.} (2018) \cite{chang2018sleepguard}           & k-NN                                   & Acc. = 71\%                \\ \cline{2-5} 

                                                     &\multirow{3}{*}{\makecell[l]{Network, Location\\and Application}} &    Wan and Lin (2016) \cite{wan2016classifying}                                       & Fuzzy Classification                   & Acc. = 96\%       \\ \cline{3-5} 
                                                     &                                                    & Chen \textit{et al.} (2018) \cite{chen2018wifi}                   & CNN                                    & Acc. = 97.7\%  \\ \cline{3-5}         
                                                    &  & Ma \textit{et al.} (2021) \cite{ma2021location}                   & 2D CNN + RNN                           & Acc. =  83\%               \\ \hline
\multirow{4}{*}{\makecell[l]{Health Parameters and \\ Body Features}} & \multirow{2}{*}{Motion}                            & Yao \textit{et al.} (2020) \cite{Yao2020MotionToBMI}              & CNN+ LSTM                              & Acc. = 94.8\%              \\ \cline{3-5} 
                                                     &                                                    & Hussain \textit{et al.} (2021) \cite{hussain2021security}         & Na\"{i}ve Bayes                            & Acc. = 71\%                \\ \cline{2-5} 
                                                     & Touchscreen                                        & Arroyo-Gallego \textit{et al.} (2017) \cite{ArroyoGallego2017}    & SVM                                    & AUROC = 88\%               \\ \cline{2-5} 
                                                     & \makecell[l]{Network, Location\\and Application}                  & Palmius \textit{et al.} (2016)\cite{palmius2016detecting}         & Linear Regression                      & Acc. = 85\%                \\ \hline
\multirow{5}{*}{Mood and Emotion}                    & \multirow{2}{*}{Motion}                            & Quiroz \textit{et al.} (2018) \cite{quiroz2018emotion}            & RF                                     & AUROC \textgreater 81\%   \\ \cline{3-5} 
                                                     &                                                    &  Neal and Canavan (2020) \cite{Neal2020Mood}                                        & RF                                     & F1-Score \textgreater 95\% \\ \cline{2-5} 
                                                     & \multirow{2}{*}{Touchscreen}                       & Gao \textit{et al.} (2012) \cite{Gao2012}                         & SVM                                    & Acc. = 69\%                \\ \cline{3-5} 
                                                     &                                                    & Shah \textit{et al.} (2015) \cite{Shah2015}                       & Lienar Regressin                       & Acc. 90.47\%               \\ \cline{2-5} 
                                                     & \makecell[l]{Network, Location\\and Application}                  & Zhang \textit{et al.} (2018) \cite{Zhang2018}         & Factor Graph                           & Acc. = 62.9\%              \\ \hline
\multirow{3}{*}{Location Tracking}                   & \multirow{2}{*}{Motion}                            &  Hua \textit{et al.} (2017) \cite{Hua2017WeCanTrack}               & Na\"{i}ve Bayes + Decision Tree         & Acc. = 92\%     \\ \cline{3-5} 
                                                     &                                                    & Nguyen \textit{et al.} (2019) \cite{Nguyen2019Location}           & DTW                                    & KL-Score = 0.057\%                   \\ \cline{2-5} 
                                                     & \makecell[l]{Network, Location\\and Application}                  & Singh \textit{et al.} (2018) \cite{singh2018ensemble}             & RF                                     & Acc. = 85.7\%              \\ \hline
\multirow{3}{*}{\makecell[l]{Keystroke Logging and \\ Text Inferring}}                 & \multirow{3}{*}{Motion}                            & Cain and Chen \textit{et al.} (2011) \cite{cai2011touchlogger}    & Gaussian Distribution                  & Acc. = 70\%           \\ \cline{3-5} 
                                                     &                                                    & Aviv \textit{et al.} (2012) \cite{Aviv2012Practicality}           & HMM                                    & Acc. = 73\%                \\ \cline{3-5} 
                                                     &                                                    &    Owusu \textit{et al.} (2012) \cite{Owusu2012ACCessory}            & Hierarchical Classifier                & Acc. = 93\%                  \\ \hline
\end{tabular}
\end{table*}

An outline of the different sensitive aspects of mobile device users that can be extracted from the different mobile device sensors, with some of the most important work in each field, is shown in Table \ref{tab:TechniquesDatasanitation}. In the remainder of this Section, examples of the personal and sensitive information extracted from the mobile device sensor data are presented, grouped in several categories depending on the nature of the extracted information and arranged by the particular data acquisition sensor.


\subsection{Demographics}
\label{subsec:demographics}

Probably the largest share of personal and sensitive information extracted from mobile user interaction data consists of attributes such as age, gender and ethnicity, which can all be ascribed to the category of demographics. 

\subsubsection{Motion Sensors}
In~\cite{Davarci2017AgeGroup} the user age range was extracted from the accelerometer data, while performing a task on the device screen. The authors exploited the \textit{k} Nearest Neighbours (\textit{k}-NN) algorithm, obtaining an accuracy of 85.3\%. Similarly, Nguyen \textit{et al.} \cite{nguyen2019kid} developed a method to distinguish an adult from a child exploiting the behavioural differences captured by the motion sensors. Based on the hypothesis that children, with smaller hands, will tend to be more shaky, they achieved an accuracy of 96\% using the Random Forest (RF) method. In \cite{Jain2016InvestigatingGender}, the gender of the users was determined with an accuracy of 76.8\% by processing with Support Vector Machines (SVM) and bagging algorithms their walking patterns acquired by a smartphone motion sensors. Meena and Saeawadekar \cite{meena2020gender} based their work on Ensemble Boosted Tree (EBT) and achieved an accuracy of 96.3\%.
The authors in \cite{Singh2019SideChannel} also focused on gender recognition from the data extracted by the accelerometer and gyroscope, obtaining an accuracy of 80\%. Ngo \textit{et al.} \cite{Ngo2019OUISIR} focused on extracting gender and age with Hidden Markov Models (HMMs). The authors organised a competition based on accelerometer and gyroscope data acquired by wearable devices, which lead to a percentage error rate of 24.23\% for gender and 5.39\% for age. With the development of deep learning techniques, it has been possible to achieve enhanced results, as in the case of Sabir \textit{et al.} \cite{sabir2019gait}, who obtained an accuracy of 94.11\% by using Long Short-Term Memory (LSTM) Recurrent Neural Networks (RNNs), a class of deep learning models particularly apt to capture temporal dependencies underlying in the data.

\subsubsection{Touchscreen}
In~\cite{Acien2019ActiveDetection}, the authors performed an analysis to identify whether the user using the device was a child or an adult based on swipe and tap gestures. For this purpose, an Active User Detection (AUD) algorithm has been used, achieving 97\% accuracy. In \cite{tolosana2021child}, a new database of children's mobile interaction was presented. The authors used touch interaction information to classify children into three groups aged 18 months to 8 years old. Nguyen \textit{et al.} \cite{nguyen2019kid} also conducted a study using RF on tap gestures to distinguish between an adult and a child, achieving an accuracy of 99\%. 
Touchscreen data has also been used to extract a person's gender. Miguel-Hurtado \textit{et al.} in \cite{miguel2016predicting} focused their work on the prediction of soft-biometrics from swipe gesture data.  They achieved 78\% accuracy rate using a decision voting scheme. In \cite{jain2019gender} the authors used gestural attributes in which the k-NN classifier recognises the gender of the user, providing a classification accuracy of 93.65\%.

\subsubsection{Network, Location and Application} 
Studies have shown a strong correlation between a user's geolocation and usage patterns and their demographics \cite{yuan2012correlating, scherrer2018travelers, almaatouq2016mobile}. For instance, in \cite{Riederer2015} the authors showed how demographic information can be inferred from geo-tagged photos on social networks. Specifically, they performed an analysis of how a person's ethnicity can be extracted from their location. They distinguished between people belonging to three different ethnicity groups with an accuracy of 72\% using a logistic regression. Also, Wu \textit{et al.} \cite{Wu2019} studied how from the spatio-temporal characteristics and geographical context extracted by GPS, they were able to obtain information on marital status and state of residence with an accuracy of 80\% based on an XGBoost algorithm. 
From the behavioural patterns found in application, Bluetooth, and Wi-Fi usage, the authors in \cite{neal2018gender} explained how it was possible to estimate the gender of a person with an accuracy of 91.8\% using RF and multinomial Na\"{i}ve Bayes.

\subsection{Activity and Behaviour}
\label{subsec:activity_behaviour}

It has been shown that a broad variety of users' behaviour or activities can be inferred from mobile device sensor data.

\subsubsection{Motion Sensors}
In ~\cite{Sun2010Activity} the authors were able to detect whether the person was stationary, walking, running, bicycling, climbing stairs, going down stairs or driving using only the accelerometer information. Their proposed approach, based on SVM, was able to achieve an accuracy of 93.2\%. Using accelerometer and gyroscope data, Anjum and Ilyas \cite{anjum2013activity} built an application where the activity that users were doing was tracked. The mobile device was in the hand, trouser pocket, shirt pocket or handbag. Using a Decision Tree classifier, they achieved an average Area Under the Receiver Operating Characteristic (AUROC) curve of over 99.0\%. 
In~\cite{thomaz2015practical} the movements made by a user while eating were estimated by the accelerometer on a smartwatch. In \cite{Santani2018} Density-based Spatial Clustering of Applications (DBSCAN) algorithm was used, achieving an accuracy of 76.1\%. 
The amount of alcohol taken by users can also be extracted from the accelerometer data. In \cite{Arnold2015}, the number of drinks consumed was inferred from accelerometer data achieving an accuracy of 70\% using a RF algorithm.
Motion sensors have also been used to extract information related to sleep such as sleep states or sleep quality. In~\cite{chang2018sleepguard} accelerometer data from a smartwatch was used achieving an accuracy of 71.0\% using a k-NN classifier.

\subsubsection{Network, Location and Application} 
From GPS data, the authors in \cite{wan2016classifying} determined whether the user is standing, walking, or using other transportation with a fuzzy classifier monitoring the speed and angle of the person obtaining a matching rate of 96\% at a five-second interval.
Wi-Fi can reveal a significant amount of information about users' activity. In \cite{chen2018wifi} the Wi-Fi Received Signal Strength Indicator (RSSI) was used on a smartphone to determine what activity users were doing. A 97.7\% accuracy rate was obtained with a Convolutional Neural Network (CNN). In \cite{ma2021location}, the authors used three neural networks on Channel State Information (CSI) measured by Wi-Fi: a 2D CNN as the recognition algorithm, a 1D CNN as the state machine, and an LSTM RNN as the reinforcement learning agent for neural architecture search. They were able to discriminate whether a person is sitting, standing, walking with an accuracy of 83\%.

\subsection{Health Parameters and Body Features}
\label{subsec:health_parameters_body_features}



\subsubsection{Motion Sensors}
The Body Mass Index (BMI) can be calculated by extracting the height and weight parameters of the person. Yao \textit{et al.} \cite{Yao2020MotionToBMI} uses a hybrid model with a CNN-LSTM architecture to estimate the continuous BMI value from the accelerometer and the gyroscope with a maximum accuracy of 94.8\%. From the BMI many health attributes can be inferred \cite{albanese2017body, dobner2018body}. 
Another parameter that can be measured from the accelerometer is stress. Garcia-Ceja \textit{et al.} \cite{garcia2018mental} achieved 71\% accuracy using similar user models and the Na\"{i}ve Bayes algorithm.

\subsubsection{Touchscreen}  
It is possible to identify whether a person has Parkinson's disease by analysing their keystroke writing pattern independently of the written content. In \cite{ArroyoGallego2017} the authors used a SVM algorithm achieving an Area Under the Curve (AUC) of 0.88 on this problem. In \cite{2019_FG_Parkinson_AM}, different types of features extracted from handwriting were studied as biometrics for Parkinson disease, achieving very promising results. In \cite{Bevan2016} the authors showed how people with longer thumbs perform swipe gestures in less time.

\subsubsection{Network, Location and Application} 
Anonymised geographic locations were collected to identify Bipolar Disorder (BD). In~\cite{palmius2016detecting}, the authors used a linear regression algorithm and a quadratic discriminant analysis algorithm achieving an 85\% accuracy. GPS can also determine sleep disorders, showing a good ability to detect sleep-wake stages and sleep-disordered breathing disorders (SRBD) such as obstructive sleep apnea (OSA)~\cite{tal2017validation,behar2014sleepap}.

StayActive\footnote{StayActive App: http://www.aal-europe.eu/projects/stayactive/} is an application that detects stress by analysing the behaviour of the users via smartphone, using the data from the Wi-Fi, step counter, location and battery level among others. In ~\cite{kostopoulos2015m} the authors used a combination of a mathematical model and a Machine Learning (ML) approach that extract information from the sleeping pattern of the users (largest time interval that the user did not touch his/her screen), their social interaction and their physical activity to determinate the stress level.

\subsection{Mood and Emotion}
\label{subsec:mood_emotion}
The user efficiency or motivation when performing a task changes in accordance to their mood. Thus, it can be inferred from different sensors.

\subsubsection{Motion Sensors}
In~\cite{Neal2020Mood}, Neal and Canavan found a correlation between the identity of a person and the mood. In their study, the authors observed that the subjects with the least accurate identification ($<$70\%) were those with the least mood changes using a RF classifier. The walk pattern data obtained from a smartwatch accelerometer and gyroscope can be used to determine a person's mood (happy, sad or neutral). The authors in  \cite{quiroz2018emotion} determined the mood with a RF algorithm achieving a mean AUROC of 81\%.

\subsubsection{Touchscreen}
Numerous studies have shown how, from the way a user interacts with the screen of his or her mobile device, it is possible to extract their mood \cite{Cao2017, Gao2012, Hung2016, Shah2015}. Gao \textit{et al.} \cite{Gao2012} demonstrated how finger-stroke features during gameplay could automatically discriminate between four emotional states (excited, relaxed, frustrated, bored). By means of an SVM algorithm they obtained an accuracy of 69\%. In \cite{Shah2015}, the authors predicted the emotional state of a person into one of the three states: positive, negative or neutral. They achieved an accuracy of 90.47\% using a linear regression.

\subsubsection{Network, Location and Application} 
MoodExplorer \footnote{MoodExplorer App:  https://play.google.com/store/apps/details?id=com.\\examsuniverse.moodexplorer} is an app that collects data from mobile sensors such as GPS, accelerometer and Wi-Fi among others. From them, the authors in~\cite{Zhang2018} recognised the composite emotions of users through a proposed model called Graph Factor with a performance metric called exact match of 62.9\% on average.

\subsection{Location Tracking}
\label{subsec:location_tracking}
Mobile devices usually come with built-in GPS modules for the purpose of location tracking. However, even when GPS coordinates are not available explicitly, position can be inferred by other sensors.

\subsubsection{Motion Sensors}
Several studies have shown how the position of a person can be inferred from the accelerometer, gyroscope and magnetometer while he or she is walking, driving or using public transport \cite{Nguyen2019Location, Han2012ACComplice, Han2012ACComplice, Hua2017WeCanTrack}. In \cite{Nguyen2019Location} the authors compared the pre-established routes with those taken by users while using different transport modes such as walking, train, bus or taxi. They compared both routes with a Dynamic Time Warping (DTW) algorithm obtaining a Kullback-Leibler distance of 0.00057 when the journey was made by taxi. In \cite{Hua2017WeCanTrack}, it was demonstrated how, when a person uses the metro, it is possible to track them from the accelerometer data. They achieved an accuracy of 92\% when the passenger travelled through 6 stations using boosted Na\"{i}ve Bayesian and Decision Trees algorithms.

\subsubsection{Network, Location and Application} 

From the different Wi-Fi networks to which a user connects, it is also possible to determine the position of an individual. In~\cite{singh2018ensemble} the location was determined in real time in indoor places. The authors achieved an accuracy of 85.7\% using a RF algorithm.

\subsection{Keystroke Logging and Text Inferring}
\label{subsec:keystroke_logging}


\subsubsection{Motion Sensors}  
Touchlogger~\cite{cai2011touchlogger} was an application created to determine the region of the phone touchscreen touched by the user, based on the device micro-movements captured by the accelerometer and the gyroscope. The screen was divided into 10 regions and with the help of a probability density function for a Gaussian distribution an accuracy of 70\% was obtained. Based on this result, it could be possible to identify the text that the user is writing. In this task, Owusu \textit{et al.} \cite{Owusu2012ACCessory} obtained an accuracy of 93\% using a hierarchical classification scheme. Similarly, in~\cite{Aviv2012Practicality}, with a controlled environment, the authors were able to identify the PIN entered 43\% of the times and the pattern 73\% of the time by means of logistic regression and HMM.



\section{Privacy Metrics for Sensitive Data}
\label{sec:privacy_metrics}

\begin{table}[t]\caption{Some of the most common privacy metrics grouped by the property measured. ADE - Adversary's Estimate: generally a posterior probability distribution. ADR - Adversary's Resources: computational power, time, etc. PAR - Parameters: for configuring privacy metrics. PK - Prior Knowledge: generally a prior probability distribution. TO - True Outcome: also known as ground truth, it can be used to evaluate the ADE.} 
\centering
\label{metrics_table}
\begin{tabular}{| l | l | l |}
\hline
\textbf{Property} & \textbf{Metric} & {\makecell[l]{\textbf{Input} \\ \textbf{Data}}} \\
\hline
\hline
\multicolumn{1}{|l|}{\multirow{8}{*}{Anonymity}} 
& \textit{k}-Anonymity \cite{Sweeney2002KAnonymity} & PAR \\
& \textit{m}-Invariance \cite{xiao2007m} & PAR   \\
& \textit{($\alpha$, k)}-Anonymity \cite{wong2006alpha} & PAR \\
& \textit{$\ell$}-Diversity \cite{Machanavajjhala2007LDiversity} & PAR   \\
& \textit{t}-Closeness \cite{Li2007TCloseness} & PAR, TO \\
& Stochastic \textit{t}-closeness \cite{domingo2015t} & PAR, TO  \\
& \textit{(c, t)}-Isolation \cite{chawla2005toward} & \makecell[l]{ADE, PAR,\\ TO}  \\
& \textit{(k, e)}-Anonymity \cite{zhang2007aggregate} & PAR  \\
\hline

\multicolumn{1}{|l|}{\multirow{5}{*}{\makecell[l]{Differential \\ Privacy}}} 
& \textit{(d-$\chi$)}-Privacy \cite{chatzikokolakis2013broadening} & PAR, TO \\
& Joint Differential Privacy \cite{kearns2014mechanism} & PAR, TO  \\
& Geo-indistinguishability \cite{andres2013geo} &  PAR, TO   \\
& Computational Differential Privacy \cite{mironov2009computational} & \makecell[l]{ADE, ADR,\\ PAR, TO}  \\
& Information Privacy \cite{du2012privacy} & ADE, PAR \\
\hline

\multicolumn{1}{|l|}{\multirow{6}{*}{Entropy}} 
& Entropy \cite{Shannon1948mathematical} & ADE  \\
& Cross-Entropy \cite{merugu2003privacy} & ADE, TO \\
& Cumulative Entropy \cite{freudiger2007mix} & ADE  \\
& Inherent Privacy \cite{agrawal2001design} &  ADE, TO \\
& Mutual Information \cite{lin2002using} & ADE, TO  \\
& Conditional Privacy Loss \cite{agrawal2001design} & ADE, TO  \\
\hline

\multicolumn{1}{|l|}{\multirow{4}{*}{\makecell[l]{Success \\ Probability}}} 
& Privacy Breach \cite{evfimievski2004privacy} & ADE, TO  \\
& \textit{(d-$\gamma$)}-Privacy \cite{rastogi2007boundary} & ADE, TO  \\
& \textit{($\delta$)}-Presence \cite{nergiz2007hiding} & ADE, TO  \\
& Hiding Failure \cite{Oliveira2002PrivacyPreserving} & ADE, TO  \\
\hline

\multicolumn{1}{|l|}{\multirow{1}{*}{Error}} 
& Euclidean Distance \cite{shokri2011quantifying} &  ADE, TO \\
\hline

\multicolumn{1}{|l|}{\multirow{4}{*}{Accuracy}} 
& Confidence Interval Width \cite{agrawal2000privacy} & ADE, PAR  \\
& \textit{(t-$\delta$)}-Privacy Violation \cite{kantarcioǧlu2004data} & \makecell[l]{ADE, PAR,\\ PK, TO} \\
& Size of Uncertainty Region \cite{cheng2006preserving} &  ADE \\
& Customisable Accuracy \cite{ardagna2007location} & PAR \\
\hline

\multicolumn{1}{|l|}{\multirow{2}{*}{Time}} 
& Maximum Tracking Time \cite{sampigethaya2005caravan} & ADE \\
& Mean Time to Confusion \cite{hoh2007preserving} & ADE, PAR \\
\hline

\end{tabular}
\end{table}

All privacy protection methods work by modifying the original data in order to deprive it of user sensitive information. For instance, the modified data should only reveal allowed attributes (e.g., gender) in order to maintain some data utility, in terms of available information, while other attributes (e.g., ethnicity) are suppressed. The degree of privacy achieved is typically related to the extent of data modification; however, the utility of the resulting dataset can be significantly impacted \cite{garfinkel2015identification}. 
\par In order to evaluate the effectiveness of privacy protection approaches, the degree of privacy protection achieved, as well as the residual data utility after data modification, should be quantified. The former task can be achieved through specific privacy metrics, whereas the latter can be expressed in terms of reduction of  traditional performance metrics such as accuracy or Equal Error Rate (EER).
\par User sensitive data acquired through mobile interaction is very heterogeneous and can be \textit{structured}, as in the case of high-level health data, network, location and application data, or \textit{unstructured}, i.e. motion, position, environmental, touchscreen and low-level health data. Consequently, different metrics are required depending on the specific application scenario. In this context, we will consider data after having undergone modifications in order to suppress or alter specific sensitive attributes, while retaining utility for analysis and extraction of non-sensitive information. 
\par In our discussion, privacy metrics will be classified based on their output, in other words, depends on the characteristics of the data that are measured with a specific metric. There is no specific metric that can be applied to every characteristic, so many studies use their own metrics. Table \ref{metrics_table} shows the metrics considered in our discussion and input data needed for the specific metric computation, grouped by the property measured. According to this criterion, some of the most relevant privacy metrics in the context of data acquired through mobile interaction can be grouped as follows \cite{wagner2018technical}:

\paragraph{Anonymity-based metrics} these metrics stem from the idea of \textit{k}-Anonymity \cite{Sweeney2002KAnonymity}, defined as the property of a dataset ensuring that in case of release, based on an individual's disclosed information, it is not possible to distinguish them from at least $k-1$ individuals whose information has also been disclosed. This is achieved by grouping subject data into equivalence classes with at least \textit{k} individuals, indistinguishable with respect to their sensitive attributes. \textit{k}-Anonymity is independent of the information extraction technique and it quantifies the degree of privacy exclusively considering the disclosed data. It is useful to express the degree of similarity between datasets, namely the original one and the sanitised one, or it can be applied to samples within a single dataset. However, several studies have reported some limitations of \textit{k}-anonymity, which have led to the development of new metrics based on the original, aiming to overcome some of its issues by imposing additional requirements. For instance, \textit{m}-invariance \cite{xiao2007m} modifies \textit{k}-anonymity to allow for multiple, different releases of the same dataset. \textit{($\alpha$,k)}-Anonymity \cite{wong2006alpha} imposes a predetermined maximum occurrence frequency for sensitive attributes within a class to protect against attribute disclosure. \textit{$\ell$}-diversity \cite{Machanavajjhala2007LDiversity} was developed to prevent linkage attacks by specifying the minimum diversity within an equivalence class of sensitive information, namely at least $\ell$
well-represented different sensitive values. 
For a skewed distribution of sensitive attributes, \textit{t}-closeness \cite{Li2007TCloseness} and stochastic \textit{t}-closeness \cite{domingo2015t} were introduced, starting from the idea that the distribution of sensitive values in any equivalence class must be close to their distribution in the entire dataset. Consequently, knowledge of the original distribution is needed to compute this metric. Similarly, starting from the original data distribution \textit{(c,t)}-isolation \cite{chawla2005toward} indicates the number of data samples present in the proximity of a sample predicted from the transformed data. Depending on the semantic distance between sensitive user records, such as in the case of numerical values, (\textit{k,e})-anonymity \cite{zhang2007aggregate} requires the range of sensitive attributes in any equivalence class to be greater than a predetermined safe value. Despite the highlighted shortcomings, \textit{k}-anonymity and the derived metrics are still largely employed today in a broad variety of different privacy contexts, but mainly for low-dimensional structured data \cite{aggarwal2005anonymity}. It has in fact been shown that \textit{k}-anonymity-based properties do not guarantee a high degree of protection in case of high-dimensional data. 



\paragraph{Differential Privacy-based metrics} differential privacy is a definition that has become popular thanks to its strong privacy statement according to which the data subject will not be affected, adversely or otherwise, by allowing their data to be used in any study or analysis, no matter what other studies, datasets, or information sources, are available~\cite{Dwork2014}. As discussed in Sec. \ref{sec:privacy_protection_methods}, differential privacy is generally achieved by adding noise to the original data. Therefore, in order to quantify differential privacy as a property of the data indicating the degree of privacy, it is a requirement to have knowledge of the original data. Differential privacy was defined in the context of databases to achieve indistinguishability between query outcomes, but thanks to its generality it has found application in different contexts for low-dimensional data, including biometrics and machine learning systems. It is in fact based on the requirement that independently of the presence of a particular data subject, the probability of the occurrence of any particular sequence of responses to queries is provided by a parameter, $\epsilon$, which can be chosen after balancing the privacy-accuracy trade-off inherent to the system. For a given computational task and a given value of $\epsilon$, there can be several differentially private algorithms, which might have different accuracy performances. As in the case of \textit{k}-anonymity, many metrics were originated from the initial definition of differential privacy, including approximate differential privacy, which has less strict privacy guarantees but is able to retain a higher utility \cite{dwork2006our}. \textit{d}-$\chi$-Privacy \cite{chatzikokolakis2013broadening} allows different measures for the distance between datasets than the Hamming distance used in the definition of differential privacy.  Joint differential privacy \cite{kearns2014mechanism} applies to systems where a data subject can be granted access to their own private data but not to others'. In the context of location privacy, geo-indistinguishability \cite{andres2013geo} is achieved by adding differential privacy-compliant noise to a geographical location within a determined distance. In contrast to previously described metrics based on differential privacy, computational differential privacy \cite{mironov2009computational} adopts a weaker adversary model, favouring accuracy. In order to adopt computational differential privacy, it is necessary to have knowledge of the posterior data distribution reconstructed from the transformed data. Similarly, information privacy \cite{du2012privacy} is met if the probability distribution of inferring sensitive data does not change due to any query output. 


\paragraph{Entropy-based metrics} in the field of information theory, entropy describes the degree of uncertainty associated to the outcome of a random variable \cite{Shannon1948mathematical}. Metrics based on entropy are generally computed from the estimated distribution of real data obtained from the sanitised data, even though additional information can be needed for a particular metric, such as the original data or some of the data transformation parameters. When attempting to estimate sensitive information from protected user data, high uncertainty generally correlates with high privacy. Nonetheless, a correct guess based on uncertain information can still occur. In \cite{merugu2003privacy}, the degree of privacy protection is quantified by cross-entropy (also referred to as likelihood) of the estimated and the true data distribution in the case of clustered data derived from the original data. A cumulative formulation of entropy was defined in \cite{freudiger2007mix} in the context of location privacy to measure how much entropy can be gathered on a route through a series of independent zones. Inherent privacy \cite{agrawal2001design} represents another example of metric derived from the definition of entropy, considering the number of possible different outcomes given a number of binary guesses. Mutual information and conditional privacy loss \cite{lin2002using, agrawal2001design} are also metrics based on entropy. The former provides a measure of the quantity of information common to two random variables and it can be computed as the difference between entropy and conditional entropy, also known as equivocation, which is useful to compute the amount of information needed to describe a random variable, assuming knowledge of another variable belonging to the same dataset. The latter property is built on similar premises, but it considers the ratio between true data distribution and the amount of information provided by another variable revealed.

\paragraph{Success Probability-based metrics} metrics in this category do not take into account properties of the data but only the outcome of sensitive information extraction attempts, as low success probabilities indicate high privacy. However, even if this trend is observable considering the entire dataset, single users' private data could still be compromised. 
In \cite{evfimievski2004privacy}, based on the original and estimated data, a privacy breach is defined as the event of the reconstructed probability of an attribute, given its true probability, being higher than a fixed threshold, whereas in \cite{rastogi2007boundary}, this idea was extended by \textit{(d,$\gamma$)}-privacy, in which additional bounds are introduced for the ratio between the true and reconstructed probabilities. In contrast, $\delta$-presence \cite{nergiz2007hiding} evaluates the probability of inferring that an individual is part of some published data, assuming that an external database containing all individuals in the published data is available. Hiding Failure (HF) \cite{Oliveira2002PrivacyPreserving} is a data similarity metric used to detect sensitive patterns. This metric is computed as the ratio between the sensitive patterns found in the sanitised data set and those found in the original data set. If HF is equal to zero, it means all the patterns are well hidden. 

\paragraph{Error-based metrics} these metrics measure the effectiveness of the sensitive information extraction process, for example, using the distance between the original data and the estimate. A lack of privacy generally takes place in case of small estimate errors. In location privacy, the expected estimation error measures the inference correctness by computing the expected distance between the true location and the estimated location using a distance metric, such as the Euclidean distance \cite{shokri2011quantifying}.
Furthermore, with particular regard to high-dimensional, unstructured data such as the ones acquired by mobile background sensors or images, a simple but common approach to quantify privacy consists in comparing the traditional performance metrics of sensitive attribute extraction methods (i.e. accuracy) before and after the data modification process. A significant performance drop is a valid indicator of the effectiveness of a data modification technique.

\paragraph{Accuracy-based metrics} quantify the accuracy of the inference mechanism, as inaccurate estimates typically show higher privacy. 
The confidence interval width indicates the amount of privacy given the estimated interval in which the true outcome lies \cite{agrawal2000privacy}. It is expressed in percentage terms for a certain confidence level.
\textit{(t, $\delta$)} privacy violation \cite{kantarcioǧlu2004data} provides information whether the release of a classifier for public data is a privacy threat, depending on how many training samples are available to the adversary algorithm. Training samples link public data to sensitive data for some individuals, and privacy is violated when it is possible to infer sensitive information from public data for individuals who are not in the training samples. 
In location privacy, the size of the uncertainty region denotes the minimal size of the region to which it is possible to narrow down the position of a target user, while the coverage of sensitive region evaluates how a user’s sensitive regions overlap with the uncertainty region \cite{cheng2006preserving}.
A different approach was proposed in \cite{ardagna2007location}. In this work, data subjects are given the possibility to customise the accuracy of the region they are in when submitting it to an internet service. The accuracy of the obfuscated region can therefore be seen as an indicator of privacy.

\paragraph{Time-based metrics} time-based metrics measure the time that elapses before sensitive information can be extracted. For instance, in location tracking, to evaluate a given privacy protection method, it can be useful to measure for how long it is possible to breach privacy by successfully tracking the user, by computing the maximum tracking time \cite{sampigethaya2005caravan} or the mean time to confusion \cite{hoh2007preserving}.

\section{Privacy Protection Methods for Sensitive Data}
\label{sec:privacy_protection_methods}
Given the amount of personal and sensitive information that can be extracted from mobile device sensors, it is necessary to apply a series of techniques to protect the data, as specified in the GDPR. The data should be used for its primary purpose, consented by the user, and it should not be possible to obtain additional information from the re-purposed data. Privacy protection methods aim to decrease the effectiveness of information extraction tools by transforming data with regard to specific sensitive attributes, while preserving the utility of the data for the original application scenario. 
In the remainder of this section, the discussed methods are grouped according to the type of input data they work on: \textit{(i)} \textit{traditional data modification techniques} work well with structured data, as most of them were developed for the purpose of disclosing sanitised datasets and their application fulfils the requirements of some of the properties discussed above, thus guaranteeing a certain degree of privacy; \textit{(ii)} \textit{machine learning-based data modification techniques}, which are more apt in the case of complex \textit{unstructured} data, as the relationship between privacy gains and information loss changes completely for high-dimensional, highly correlated unstructured data like images, audio signals and time sequence signals provided by background sensors in mobile devices \cite{WIERINGA2021915, liangyuan2018}. An overview of the different privacy protection methods can be found in Table \ref{tab:PrivacyProtectionMethods}.

\begin{table*}[htbp!]
\centering
\caption{Comparison of different state-of-the-art Privacy Protection Methods for Sensitive Data. AE- Autoencoder, SGD- Stochastic Gradient Descent, CNN- Convolutional Neural Network, GAN- Generative Adversarial Network, SAN- Semi-Adversarial Network,  FAR- False Acceptance Rate, TASR- Task Assigment Sucess Rate, HF- Hiding Failure, Acc- Accuracy,  SA- Sensitive Attribute, AD- Attribute Disclosure, IVE- Incremental Variable Eliminator, COCR- Correct Overall Classification Rate, LFW- Labeled Faces in the Wild}
\label{tab:PrivacyProtectionMethods}
\begin{tabular}{|l|l|l|l|l|l|}
\hline
\multicolumn{6}{|c|}{\multirow{2}{*}{\textbf{Traditional Methods}}}                                                                                                                                                                                                                                                                                                                                                                                                                                                                                                                                                                                                                                                                                                                                              \\ 

\multicolumn{6}{|c|}{}                                                                                                                                                                                                                                                                                                                                                                                                                                                                                                                                                                                                                                                                                                                                                                                           \\ \hline \hline
\textbf{Method/ Classifier}                                                   & \textbf{Field}                                                                               & \textbf{Sensitive Data Protected} & \textbf{Study}                                                                                                                                   & \textbf{Best Performance}                                                                                                                                                                                                                                                                                         & \textbf{Database}                                                                                                      \\ \hline
\multirow{2}{*}{Data Perturbation}                                            & \begin{tabular}[c]{@{}l@{}}Fingerprint\\ Faces Images\end{tabular}                           & Demographics                      & \begin{tabular}[c]{@{}l@{}}Sadhya \textit{et al.} \\ (2016) \cite{sadhya2016privacy}\end{tabular}              & \begin{tabular}[c]{@{}l@{}}0.45$\%$ probability of success\\ @ FAR = 10$\%$\end{tabular}                                                                                                                                                                                                                              & \begin{tabular}[c]{@{}l@{}}VC2002-DB1 Database\\ AR Face Database\end{tabular}                                         \\ \cline{2-6} 
                                                                              & Location Data                                                                                & Location Tracking                 & \begin{tabular}[c]{@{}l@{}}Yang \textit{et al.} \\ (2018) \cite{yang2018density}\end{tabular}                  & TASR $\approx$ 80 $\%$                                                                                                                                                                                                                                                                                 & \begin{tabular}[c]{@{}l@{}}SimpleGeo \\ Places Database\\ Yelp Database\end{tabular}                                      \\ \hline
Data Blocking                                                                 & Weather Parameters                                                                           & Health Parameters                 & \begin{tabular}[c]{@{}l@{}}Parmar \textit{et al.} \\ (2011) \cite{parmar2011blocking}\end{tabular}             & HF = 0/3 attribute disclosure                                                                                                                                                                                                                                                                                     & \begin{tabular}[c]{@{}l@{}}UCI Repository: \\ Weather Dataset\end{tabular}                                                                                           \\ \hline
\begin{tabular}[c]{@{}l@{}}Data Aggregation\\ or Merging\end{tabular}         & Physiologic Signals                                                                          & Health Parameters                 & \begin{tabular}[c]{@{}l@{}}Ren \textit{et al.} \\ (2013) \cite{ren2013sensitive}\end{tabular}                  & \multicolumn{1}{c|}{-}                                                                                                                                                                                                                                                                                            & \begin{tabular}[c]{@{}l@{}}MIT-BIH \\ Polysomnographic \\ Database\end{tabular}                                                                                    \\ \hline
Data Swapping                                                                 & Personal Attributes                                                                          & Health Parameters                 & \begin{tabular}[c]{@{}l@{}}Hasan \textit{et al.} \\ (2016) \cite{hasan2016effective}\end{tabular}              & l-Diversity = 0 attribute disclosure                                                                                                                                                                                                                                                                              & \begin{tabular}[c]{@{}l@{}}UCI Repository: \\ Synthetic Dataset\\ Adult Dataset\end{tabular}                         \\ \hline
Data Sampling                                                                 & Personal Attributes                                                                          & Health Parameters                 & \begin{tabular}[c]{@{}l@{}}Liu \textit{et al.}\\  (2019) \cite{liu2019novel}\end{tabular}                      & l-Diversity $\approx$ 0.15 error                                                                                                                                                                                                                                                                     & \begin{tabular}[c]{@{}l@{}}UCI Repository: \\ Adult Dataset\end{tabular}                                                                                           \\ \hline
\multicolumn{6}{|c|}{\multirow{2}{*}{\textbf{Machine Learning-based Methods}}}                                                                                                                                                                                                                                                                                                                                                                                                                                                                                                                                                                                                                                                                                                                                              \\ 

\multicolumn{6}{|c|}{}                                                                                                                                                                                                                                                                                                                                                                                                                                                                                                                                                                                                                                                                                                                                                                                           \\ \hline \hline

\textbf{Method/ Classifier}                                                   & \textbf{Field}                                                                               & \textbf{Sensitive Data Protected} & \textbf{Study}                                                                                                                                   & \textbf{Best Performance}                                                                                                                                                                                                                                                                                         & \textbf{Database}                                                                                                      \\ \hline
\multicolumn{6}{|c|}{Data Level Methods}                                                                                                                                                                                                                                                                                                                                                                                                                                                                                                                                                                                                                                                                                                                                                                         \\ \hline
\begin{tabular}[c]{@{}l@{}}Differential Privacy-based\\ AE\end{tabular}       & \begin{tabular}[c]{@{}l@{}}Activity Signals,\\ Biomarkers,\\ Biometric Measures\end{tabular} & Health Parameters                 & \begin{tabular}[c]{@{}l@{}}Phan \textit{et al.} \\ (2016) \cite{PhanWangWuDou2016}\end{tabular}                & Acc. Privacy $\approx$ 85 $\%$                                                                                                                                                                                                                                                                         & Own Database                                                                                                           \\ \hline
SGD sanititation                                                              & Language Modeling                                                                            & Text Inferring                    & \begin{tabular}[c]{@{}l@{}}McMahan \textit{et al.}\\  (2018) \cite{mcmahan2018learning}\end{tabular}           & \begin{tabular}[c]{@{}l@{}}-0.13$\%$ in accuracy with\\  (4.6e10$^{-9}$)-differential privacy\end{tabular}                                                                                                                                                                                                          & Reddit Dataset                                                                                                         \\ \hline
\multirow{2}{*}{Siamese CNN}                                                  & Face Images                                                                                  & Identity                          & \multirow{2}{*}{\begin{tabular}[c]{@{}l@{}}Osia \textit{et al.} \\ (2019) \cite{Osia2020AHybrid}\end{tabular}} & \begin{tabular}[c]{@{}l@{}}EER before $\approx$ 1 $\%$\\ EER after $\approx$ 28 $\%$\end{tabular}                                                                                                                                                                                           & \begin{tabular}[c]{@{}l@{}}IMDB-Wiki +\\  LFW Datasets\end{tabular}                                                                                                \\ \cline{2-3} \cline{5-6} 
                                                                              & Activity Signals                                                                             & Demographics                      &                                                                                                                                                  & \begin{tabular}[c]{@{}l@{}}EER before $\approx$ 22 $\%$\\ EER after $\approx$ 36 $\%$\end{tabular}                                                                                                                                                                                          & MotionSense Dataset                                                                                                    \\ \hline
Siamese CNN                                                                   & Activity Signals                                                                             & Demographics                      & \begin{tabular}[c]{@{}l@{}}Garofalo \textit{et al.}\\  (2019) \cite{garofalo2019data}\end{tabular}             & \begin{tabular}[c]{@{}l@{}}F1-score SA before = 72.58 $\%$\\ F1-score SA after = 52.99 $\%$\end{tabular}                                                                                                                                                                                                              & OU-ISIR Database                                                                                                       \\ \hline
GAN                                                                           & Activity Signals                                                                             & Demographics                      & \begin{tabular}[c]{@{}l@{}}Ngueveu \textit{et al.} \\ (2020) \cite{Ngueveu2020DYSAN}\end{tabular}              & \begin{tabular}[c]{@{}l@{}}Acc. SA before = 98.5 $\%$\\ Acc. SA after = 61.0$\%$\\ \\ Acc. SA before = 98.5$\%$\\ Acc. SA after = 57.0$\%$\end{tabular}                                                                                                                                                                   & \begin{tabular}[c]{@{}l@{}}MotionSense Dataset\\ \\ \\ MobiAct Dataset\end{tabular}                                    \\ \hline
\multirow{2}{*}{SAN}                                                          & Face Images                                                                                  & Demographics                      & \begin{tabular}[c]{@{}l@{}}Mirjalili \textit{et al.} \\ (2018) \cite{mirjalili2018semi}\end{tabular}           & \begin{tabular}[c]{@{}l@{}}Error Rate SA before = 19.7 $\%$\\ Error Rate SA after = 39.3 $\%$\\ \\ Error Rate SA before = 8.0 $\%$\\ Error Rate SA after = 39.2 $\%$\\ \\ Error Rate SA before = 33.4 $\%$\\ Error Rate SA after = 72.5 $\%$\\ \\ Error Rate SA before = 16.9 $\%$\\ Error Rate SA after = 53.8 $\%$\end{tabular} & \begin{tabular}[c]{@{}l@{}}CelebA Dataset\\ \\ \\ MORPH Dataset\\ \\ \\ MUCT Dataset\\ \\ \\ RaFC Dataset\end{tabular} \\ \cline{2-6} 
                                                                              & Face Images                                                                                  & Demographics                      & \begin{tabular}[c]{@{}l@{}}Mirjalili \textit{et al.}\\  (2020) \cite{mirjalili2020privacynet}\end{tabular}     & \begin{tabular}[c]{@{}l@{}}EER SA before $\approx$ 1 $$\%$$ \\ EER SA after = 20 $$\%$$\\ \\ EER SA after = 20 $$\%$$ \\ \\ EER SA after = 10 $$\%$$ \\ \\ EER SA after = 10 $$\%$$\end{tabular}                                                                                                                            & \begin{tabular}[c]{@{}l@{}}CelebA Dataset\\ \\ \\ UTK-face Dataset\\ \\ MORPH Dataset\\ \\ MUCT Dataset\end{tabular}   \\ \hline
\multicolumn{6}{|c|}{Features Level Methods}                                                                                                                                                                                                                                                                                                                                                                                                                                                                                                                                                                                                                                                                                                                                                                     \\ \hline
Decision Tree Ensemble                                                        & Face Images                                                                                  & Demographics                      & \begin{tabular}[c]{@{}l@{}}Terhorst \textit{et al.}\\  (2019) \cite{terhorst2019suppressing}\end{tabular}      & \begin{tabular}[c]{@{}l@{}}COCR before = 94.7 $\%$\\ COCR after = 64.7 $\%$\end{tabular}                                                                                                                                                                                                                              & FERET Database                                                                                                         \\ \hline
AE                                                                            & Face Images                                                                                  & Demographics                      & \begin{tabular}[c]{@{}l@{}}Bortolato \textit{et al.}\\ (2020) \cite{bortolato2020learning}\end{tabular}        & \begin{tabular}[c]{@{}l@{}}EER SA before = 1.8 $\%$\\ EER SA after = 41.9 $\%$\\ \\ EER SA before = 4.9 $\%$\\ EER SA  after = 41.4 $\%$\\ \\ EER SA before = 14.5 $\%$\\ EER SA after = 50.2 $\%$\end{tabular}                                                                                                               & \begin{tabular}[c]{@{}l@{}}CelebA Dataset\\ \\ \\ LFW Dataset\\ \\ \\ Adience Dataset\end{tabular}                     \\ \hline
\begin{tabular}[c]{@{}l@{}}Sensitivity Detector +\\ Triplet Loss\end{tabular} & Face Images                                                                                  & Demographics                      & \begin{tabular}[c]{@{}l@{}}Morales \textit{et al.}\\  (2020) \cite{morales2020sensitivenets}\end{tabular}      & \begin{tabular}[c]{@{}l@{}}Acc. SA before = 95.1 $\%$\\ Acc. SA after = 54.6 $\%$\end{tabular}                                                                                                                                                                                                                        & DiveFace Database                                                                                                      \\ \hline
\end{tabular}
\end{table*}
\subsection{Traditional Data Modification Methods}
\label{subsec:traditional_data_modification_methods}

Traditional data modification techniques have proven to work well with structured data. According to \cite{Verykios2004state}, these methods can be divided into the following groups:
\subsubsection{Data Perturbation} it is accomplished by the alteration of an attribute value by a new value. Among traditional data perturbation approaches, randomisation techniques are based on the use of noise to mask the values of the data \cite{Aggarwal2008}. By incorporating sufficiently large noise, individual data can in fact no longer be recovered, whilst the probability distribution of the aggregate data can be recovered and used safely from a privacy protection standpoint. Noise can be added to the original values in a number of ways: additive noise \cite{mivule2013utilizing, Brand2002}, multiplicative noise \cite{kim2003multiplicative}, geometric perturbation, in which a mix of additive and multiplicative perturbations are used through a rotation matrix \cite{1565733}, nonlinear transformation~\cite{5497205, LYU201821}, data condensation \cite{10.1007/978-3-540-24741-8_12} or through a combination of the above techniques \cite{CHAMIKARA20181}. 
\par Differential privacy has been widely used in several applications such as the data perturbation technique. For instance, in~\cite{sadhya2016privacy}, differential privacy was used in a privacy-preserving framework for a recognition system based on soft biometrics, such as age, gender, height, and weight extracted from fingerprints and face images. In the context of mobile devices, differential privacy has also been applied for providing rigorous protection of worker locations in a company centralised server crowdsensing application \cite{yang2018density}. 
\subsubsection{Data blocking} the replacement of an existing attribute value with a predetermined value to indicate the data suppression (it could be ``?", ``0" or ``x" in the case of one-character values) \cite{karakasidis2015blocking, parmar2011blocking}. 
\subsubsection{Data Aggregation or Merging} the combination of values in a coarser category \cite{li2012efficient, ren2013sensitive}. 
\subsubsection{Data Swapping} refers to interchanging values of
individual records \cite{hasan2016effective}.
\subsubsection{Data Sampling} the releasing data of a sample of the population \cite{chaudhuri2006when, liu2019novel}.
\par Such strategies have found a large number of different implementations for structured data and are often adopted by governmental or statistical agencies. Many are available in libraries under open-source license, like ARX\footnote{Available at \url{https://arx.deidentifier.org}.} or the R-package sdcMicro \cite{JSSv067i04, WIERINGA2021915}. However, a critical aspect of these modification techniques is often scalability, i.e. there is a significant performance drop as the number of the dimensions of the dataset increases; in addition, the computational overhead will increase exponentially with respect to the number of attributes and number of instances. These limitations of the traditional data modification methods are commonly grouped under the label of ``curse of dimensionality” \cite{koppen2000curse}. 

\subsection{Machine Learning-based Data Modification Methods}
\label{subsec:ml_data_modification_methods}

In addition to the goal of information extraction as discussed in Sec. \ref{sec:privacy_sensitive_data}, considering its potential in big data processing \cite{qiu2016survey}, machine learning approaches have in turn been investigated for the purpose of perturbing the data in the attempt to overcome the limitations of traditional modification techniques.
Within these algorithms, a subdivision into two groups can be made of those that operate at the \textit{i)} data level and those that operate at the \textit{ii)} feature level, depending on the input data. In this section we present a brief summary of the most competitive techniques of the two groups according to \cite{bortolato2020learning}.

\subsubsection{Data level methods}
Algorithms that operate at the data level have raw data as input. Within the algorithm itself they are processed and the output is a transformed dataset containing the protected sensitive data.
\par Among privacy protection solutions adopted to protect sensitive data in the context of machine learning models, differential privacy-based mechanisms are popular in the literature. In \cite{PhanWangWuDou2016}, a differentially private model implementation based on perturbing the objective functions was proposed for deep Autoencoders (AE) for human behaviour prediction in a health social network. Such a method, can be applied to each layer of the network. Similarly, the idea of sanitising the gradient in Stochastic Gradient Descent (SGD) was introduced in \cite{Abadi2016deep} for CNN, and for complex sequence models for next-word prediction in \cite{mcmahan2018learning}. Differential privacy has also been implemented in dedicated Tensorflow\footnote{Available at \url{https://github.com/tensorflow/privacy}.} and PyTorch\footnote{Available at \url{https://github.com/pytorch/opacus}.} libraries. Generally, however, at a modest privacy budget differentially private mechanisms come with a cost in software complexity, training efficiency, and model quality \cite{tramer2021differentially}.
\par Another particular case is the Siamese CNN. Using a convolutional architecture, it has two different input vectors with which it works with the same weights to acquire comparable output vectors. Osia \textit{et al.} \cite{Osia2020AHybrid} used this architecture both in the field of facial images, to protect the identity of the person, and in the field of activity recognition to protect the gender of the user. The authors in \cite{garofalo2019data} also used a Siamese CNN. In this case their work focused solely on activity recognition while protecting demographic information.
\par Also, Generative Adversarial Networks (GANs) are among the most popular techniques considered for this purpose in the literature. GANs are unsupervised methods that exploit two adversarial subnetworks (the \textit{generator} and the \textit{discriminator}), and are able to learn well, in a competitive manner, the statistical structure of high dimensional signals. A GAN-based approach called DySan was developed in~\cite{Ngueveu2020DYSAN} for data sanitisation, in the context of a mobile application for physical activity monitoring through the accelerometer and the gyroscope data. Before sending the data to a server hosted on the cloud, gender inferences are prevented by distorting the data while limiting the loss of accuracy on physical activity monitoring. 
\par A similar approach for privacy protection is based on semi-adversarial networks. Semi-adversarial networks are different from typical GANs in the fact that, in addition to the generator subnetwork, they include two independent discriminator classifiers rather than one. A semi-adversarial configuration was proposed by Mirjalili \textit{et al.} \cite{mirjalili2018semi} for the purpose of image data perturbation. Based on the feedback of two classifiers, where one acts as an adversary of the other, this model was able to privatise gender while maintaining the same accuracy in face recognition. The authors extended their work in~\cite{mirjalili2020privacynet}, by including, among other things, the possibility of choosing to obfuscate specific attributes (e.g., age and race), while allowing for other types of attributes to be extracted (e.g., gender).

\subsubsection{Features level methods}

There is a second set of methods that, instead of using raw data as input, apply on the embedding representation of the data. Therefore, a pre-trained model used as features extractor is needed. After that, this set of features will be the input of the privacy method. Finally, a transformed dataset that keeps the sensitive data privatised will be the output. Terh\"ost \textit{et al.} \cite{terhorst2019suppressing} proposed an Incremental Variable Eliminations algorithm (IVE). The authors, by training a set of decision trees, obtain a measure of the importance of the variables that predict the sensitive attributes to be reduced.

\par An AE was also used by Bortolato \textit{et al.} \cite{bortolato2020learning}. The authors introduced Privacy-Enhancing Face-Representation learning Network (PFRNet), a neural network-based model that works at the level of face representations (templates) from images, aiming to achieve distinct encodings for both identity and gender in the features space.

\par Morales \textit{et al.} \cite{morales2020sensitivenets} aimed to leave out sensitive information in the decision-making process in an image-based face recognition system without a significant drop of performance by focusing on the feature space. Developed for the purpose of ensuring fairness and transparency, their systems inherently improve the privacy of the data. It works as a independent, decoupled module on top of a pre-trained model and takes as input the embeddings generated by the model. By defining and minimising its own triplet-loss function, SensitiveNets generates new representations agnostic of gender and ethnicity information, which however still retain information useful for extraction of other attributes. 

\subsection{Other Perspectives}
\label{subsec:other_perspectives}
\par Finally, it is important to highlight that in order to protect users' privacy while handling their private data, besides data modification methods, other important perspectives to be considered to comply with secure data management practices in relation to privacy include: 
\subsubsection{Template protection} is an important field of research in the area of biometrics. Templates are compact representations of users' biometric data for the purpose of storage. They are transformed into protected biometric references for security purposes. Template protection schemes should provide the following properties \cite{7192825}:
\begin{itemize}
    \item Irreversibility: it should be computationally difficult\footnote{A problem is defined computationally difficult if it cannot
    be solved using a polynomial-time algorithm.} to compute the original template from a subject’s protected biometric reference.
    \item Revocability:  it should be computationally difficult to compute the original biometric template from multiple instances of protected biometric reference derived from the same biometric trait of an individual. Biometric data is permanently associated with the data subject and it cannot be revoked and reissued if compromised, contrarily to credit cards or passwords. However, through revocable and irreversible transformations templates can be cancelable, thus mitigating the risks associated with biometric template theft \cite{7192838}.
    \item Unlinkability: it should be computationally difficult to determine whether two or more instances of protected biometric reference were obtained from the same biometric trait of a user. Unlinkability prevents cross-matching across databases.
\end{itemize}
\subsubsection{Data outsourcing} Usually mobile applications exploit cloud resources for model training and inference. Therefore, users' personal data containing sensitive information may be on the internet. If stored on the cloud, data subject privacy undergoes greater risks than being stored locally in the device \cite{SVANTESSON2010391}. Performing the training and inference tasks locally is among alternative solutions investigated. However, the computational resource constraints are much stricter \cite{ServiaRodriguez2017}. 
\par A different approach could be federated learning, a machine-learning strategy according to which models are trained on datasets distributed across multiple devices, thus preventing data leakage \cite{konecny2016federated}. However, recent attacks demonstrate that simply maintaining data locality during training processes does not provide sufficient privacy guarantees as intermediate results, if exposed, could still cause some information leakage \cite{yang2019federated}. Possible solutions to this problem are given by differential privacy mechanisms and Secure Multiparty Computation (SMC) schemes, or a combination of the two \cite{truex2019hybrid}. 
\par Finally, it should be pointed out that the considered techniques should be complemented by widely deployed encryption protocols that would guarantee data security, such as hash functions, secret-key and public-key cryptography, among others \cite{HernandezAlvarez2021}.

\section{Conclusions and Open Research Questions}
\label{sec:conclusion_and_open_directions}

\subsection{Conclusions}

As demonstrated, seemingly innocuous user data can reveal personal and sensitive information about the user, which must be protected in compliance with the GDPR. We have provided a state-of-the-art review of the different kinds of sensitive data that can be extracted by the mobile device sensor data. 
A survey of the metrics that allow a comparison of different aspects and quantify the effectiveness of the privacy protection methods was carried out for the purpose of identifying the most suitable metric for each specific application. Some of the most popular privacy protection data modification methods were also discussed, aiming to offer useful guidelines for managing the trade-off between protecting the sensitive attributes while disclosing the allowed attributes, inherent to the privacy problem.

\subsection{Open Research Questions}
Many paths of development remain to be investigated. The most relevant ones are discussed below.
\subsubsection{Protection of the Privacy of User Sensitive Data}
\paragraph{Correlation between Sensitive Attributes}
It is important to observe the correlation between the different sensitive attributes, in order to identify from which sensitive attributes it is possible to extract others. For example, the user location obtained from the mobile device Wi-Fi data can also reveal information about the activity a user is involved in.

\paragraph{Data Modification Algorithms for Privacy Protection}
As shown in Sec. \ref{sec:privacy_sensitive_data}, inferrable attributes can assume very diverse sets of values, in terms of size and number of attributes per subject. For instance, the presence of a disease or the gender of a data subject are unique to each subject and can assume a binary or limited set of values. A different scenario is given by attributes such as the age (unique attribute, but wider set of possible values), or the activity the user is involved in or their location (several possible attributes per subject, but one at a time). Depending on the formulation of the attribute output categories, at the cost of increased system complexity, it is possible to achieve a finer granularity in terms of information about the data subject, which typically relates to a higher extent of privacy invasion. Therefore, from the perspective of sensitive data protection, a possible step towards the protection of the privacy of sensitive data could be developing a system that would modify the data so that the possible sensitive attribute recognisable output categories would be fewer and coarser.

\paragraph{Ethical Implications}
The digitalisation of data storage and communications, combined with the ever-growing capacity of computers to automatically process data, has made possible to mine structures and relationships lying in the data to extract information in unprecedented ways. Among other things, the GDPR provides a definition of personal and sensitive information to safeguard the right to privacy in the digital domain, thus laying the cornerstone of an ethical usage of user data. Nonetheless, even if the sensitive information is suppressed, it would be beneficial to assess the side effects of automated processing, with regard to sensitive attributes, paying special attention to the ethical consequences this might entail. 
Therefore, even if data is collected and processed for a legitimate purpose, the results yielded might be influenced by personal and sensitive information that the models are covertly recognising and exploiting. 
For instance, in 2018, Amazon withheld their machine learning engine in charge of selecting the most suitable job applicant profiles as it was discovered that it was biased against women, downgrading resumes that included the word “women’s,” as in “women’s chess club captain” and graduates of all-women’s universities. This was due to the fact that the models utilised were trained with resumes submitted to the company over a 10-year period, which came mostly from men \cite{AmazonBias}. 
Such risks are exacerbated by the fact that, in the case of deep learning models, it is often difficult to ascertain how such information is encoded in the intermediate layers, and that the sensitive and legitimate attributes might be entangled within their representation instances. Fairness in AI is an novel, yet very active field of investigation, deeply connected with the protection of the privacy of user sensitive data.


\subsubsection{Performance of the Algorithms}
\paragraph{Robustness}
Given the ubiquity of mobile devices, the data are captured by the built-in sensors in a variety of different scenarios. Therefore, a typical requirement of mobile device computing is robustness. For instance, with regards to the recognition of sensitive information, the property of position invariance would grant a negligible impact in the performance of an algorithm due to changes in the position of the mobile device with respect to the user who is carrying. In other words, the algorithm should be able to recognise the predetermined user attributes regardless of whether the mobile device is in the front pocket, in the hand or in the backpack or whether it is performing a specific activity such as answering a call or typing.

\paragraph{Reliability of Labels}
It is important to identify whether the subjects themselves are in charge of the task of labelling the sensitive attribute. In this case it is important to ensure that a subject is able to do it in an objective way, in the case of mood recognition, for instance. Additionally, with particular regard to 3D motion sensor data in the time domain, labelling is often not straight-forward and it can be expensive and time-consuming. Improving or overcoming the labelling process is an interesting open problem for further investigation. A solution could be adopting self-supervised learning (SSL), a paradigm according to which the training of the feature extraction algorithms can take place in an unsupervised manner. 


\paragraph{Impact of Hardware Differences}

After performing a study of the different mobile device sensors, it would be interesting to evaluate how innate sensor characteristics affect the processes of sensitive data extraction and protection. This is due to the fact that not all smart device sensors have the same characteristics, i.e. full-scale values, resolution, sampling frequencies, etc.


\paragraph{Computation Time}
With regard to mobile devices, time constraints are often crucial for real-time applications, and a seamless user experience is among the main user concerns. Therefore, incorporating in the processing chain additional steps aiming to protect the privacy of the sensitive data should not impact the computation time significantly.   



\paragraph{Storage of the Algorithms}

Finally, a significant aspect is related to the storage of the algorithms. The captured user raw data may then be sent to the cloud, for training the models, as more powerful hardware resources are typically available remotely. In such way, the raw data might be exposed to greater risks related to being transmitted and stored in a server. It is therefore necessary to develop systems that achieve the desired degree of sensitive data protection, without impacting the performance of the models. Among the solutions proposed for such goals is federated learning, in combination with algorithms that would guarantee differential privacy and SMC.



\subsubsection{General metric framework}

With regard to the protection of the privacy of sensitive data, it would be desirable to create a general metric framework that can be applied to any set of protected data and indicate with certainty the degree of protection through a score, encompassing which attributes are being protected and how many classes are being used to differentiate an attribute. Based on this, a standardised set of limit values should be established in order to indicate the point at which sensitive data is considered fully protected. In such way, protected data could be freely processed for extraction of information without putting at stake the privacy of users' sensitive data.

\section*{Acknowledgments}
This work has been supported by project PriMa (MSCA-ITN-2019-860315).


\printbibliography
\vskip 0pt plus -1fil

\begin{IEEEbiography}[{\includegraphics[width=1in,height=1.25in,clip,keepaspectratio]{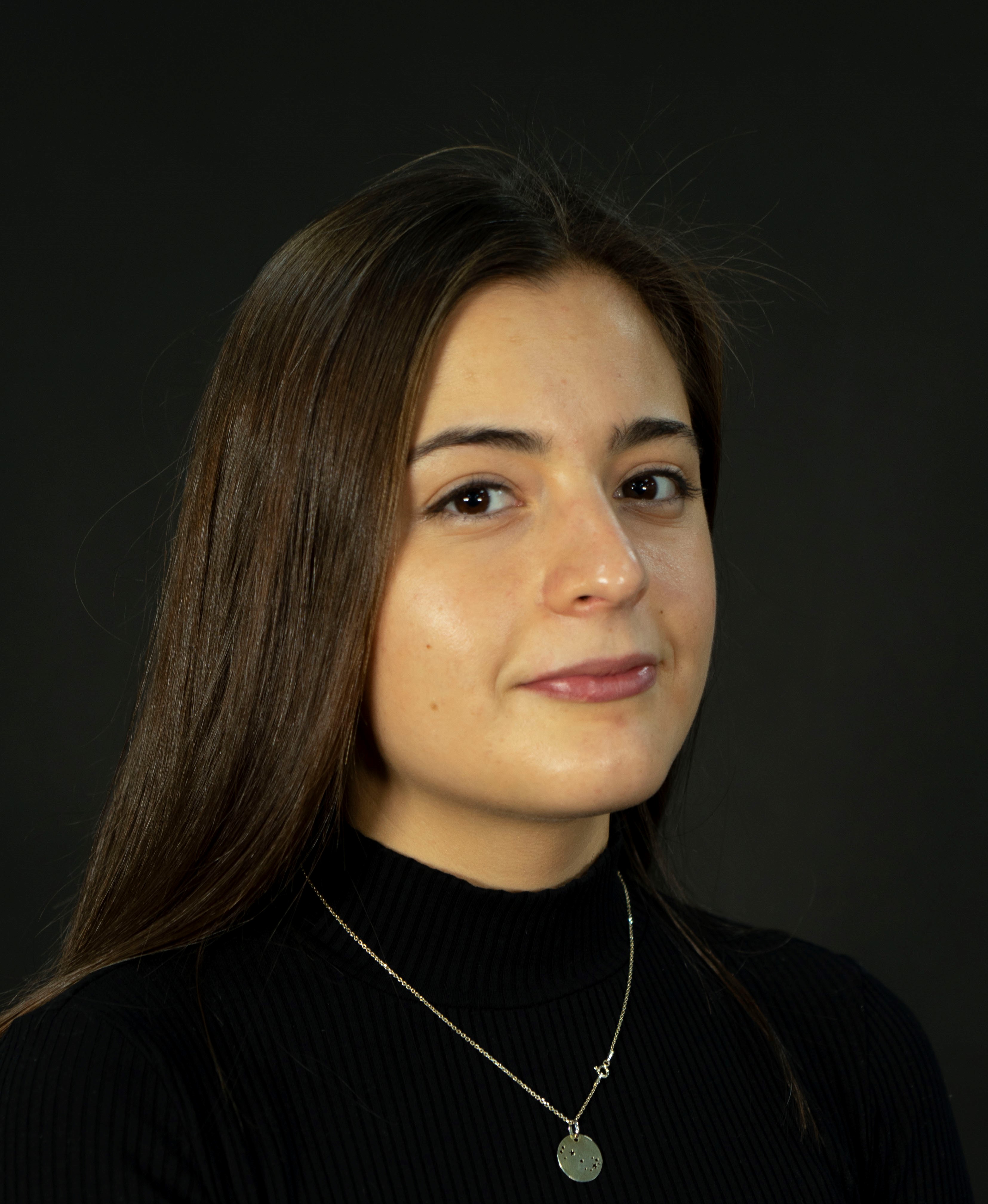}}]{Paula Delgado-Santos}
 received the M.Sc. degree in Telecommunications Engineering from Universidad Autonoma de Madrid, Spain, in 2020. At the same time, she was working in a scholarship of IBM. In 2019/2020 she was working at a Swiss University, HEIG-VD, as a Data Scientist. In 2020 she started her PhD with a Marie Curie Fellowship within the PriMa (Privacy Matters) EU project at University of Kent, United Kingdom. Her research interests include signal and image processing, pattern recognition, machine learning, biometrics and data protection.
\end{IEEEbiography}
\vskip 0pt plus -1fil

\begin{IEEEbiography}[{\includegraphics[width=1in,height=1.25in,clip,keepaspectratio]{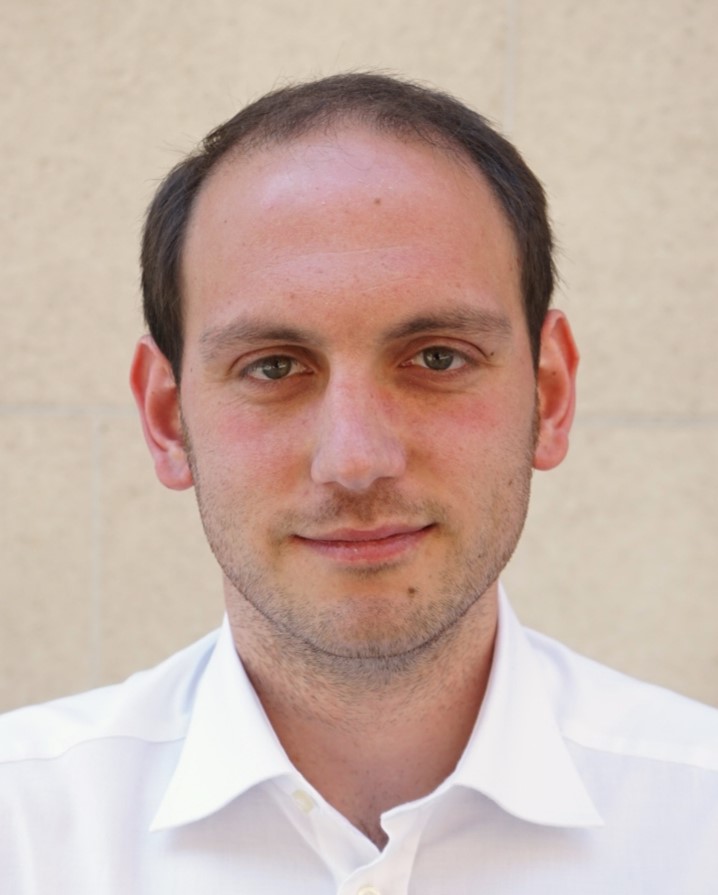}}]{Giuseppe Stragapede}
 received his M.Sc. degree in Electronic Engineering from Politecnico di Bari, Italy, in 2019. After one year as a computer vision engineer in the industry, in 2020 he started his PhD with a Marie Curie Fellowship within the PriMa (Privacy Matters) EU project in the Biometrics and Data Pattern Analytics - BiDA Lab, at the Universidad Autonoma de Madrid, Spain. His research interests include signal processing, machine learning, biometrics and data protection.
\end{IEEEbiography}
\vskip 0pt plus -1fil

\begin{IEEEbiography}[{\includegraphics[width=1in,height=1.25in,clip,keepaspectratio]{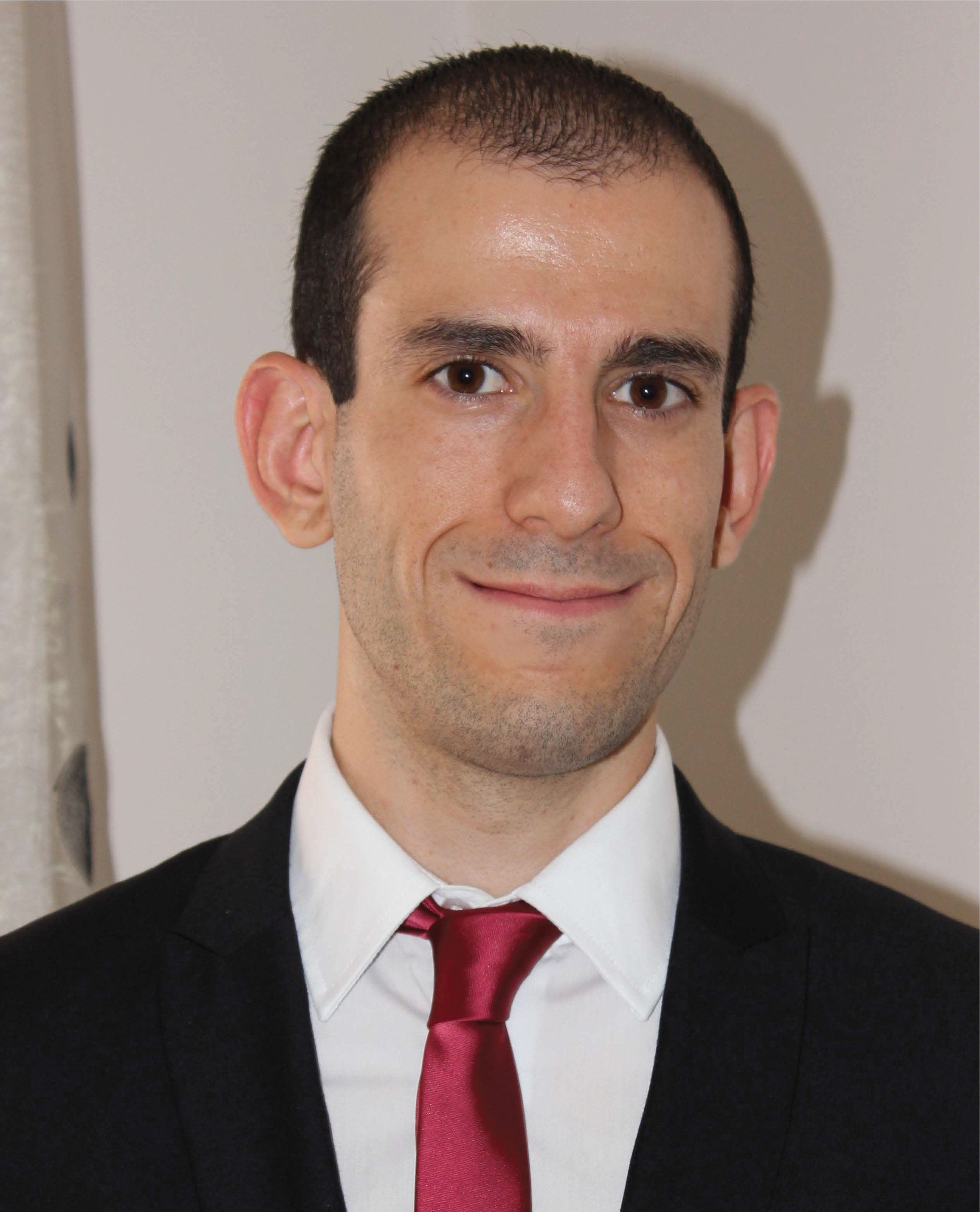}}]{Ruben Tolosana}
 received the M.Sc. degree in Telecommunication Engineering, and his Ph.D. degree in Computer and Telecommunication Engineering, from Universidad Autonoma de Madrid, in 2014 and 2019, respectively. In 2014, he joined the Biometrics and Data Pattern Analytics - BiDA Lab at the Universidad Autonoma de Madrid, where he is currently collaborating as a PostDoctoral researcher. Since then, Ruben has been granted with several awards such as the FPU research fellowship from Spanish MECD (2015), and the European Biometrics Industry Award (2018). His research interests are mainly focused on signal and image processing, pattern recognition, and machine learning, particularly in the areas of DeepFakes, HCI, and Biometrics. He is author of several publications and also collaborates as a reviewer in high-impact conferences (WACV, ICPR, ICDAR, IJCB, etc.) and journals (IEEE TPAMI, TCYB, TIFS, TIP, ACM CSUR, etc.). Finally, he is also actively involved in several National and European projects.  
\end{IEEEbiography}
\vskip 0pt plus -1fil

\begin{IEEEbiography}[{\includegraphics[width=1in,height=1.25in,clip,keepaspectratio]{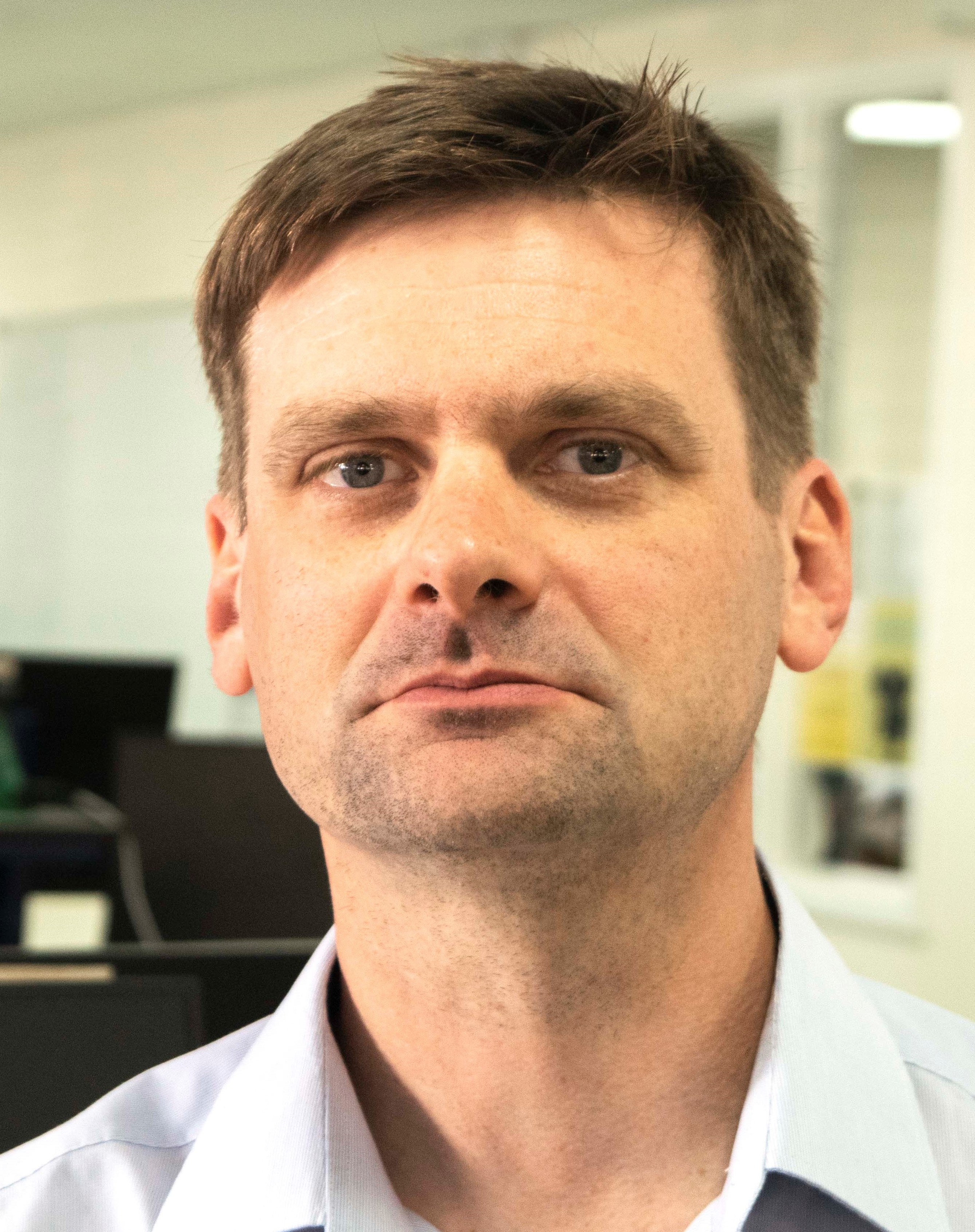}}]{Richard Guest}
 obtained his PhD in 2000. He is Professor of Biometric Systems Engineering and Head of the School of Engineering at the University of Kent. His research interests lie broadly within image processing and pattern recognition, specializing in biometric and forensic systems, particularly in the areas of image and behavioral information analysis, standardization and mobile systems.
\end{IEEEbiography}
\vskip 0pt plus -1fil
\begin{IEEEbiography}[{\includegraphics[width=1in,height=1.25in,clip]{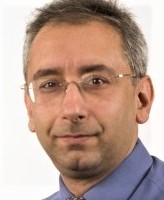}}]{Farzin Derazi}
 received the B.A. degree in Engineering Science and Economics from the University of Oxford, U.K., in 1981, the M.Sc. degree in Communications Engineering from Imperial College, U.K., in 1982, and the Ph.D. degree in Electronic Engineering from the 
University of Wales, Swansea, U.K., in 1988. He is currently with the School of Engineering and Digital Arts, University of Kent, Canterbury, U.K., where he is the Emeritus Professor of Information Engineering. His current research interests include the fields of pattern recognition and signal processing and their application in security and healthcare.
\end{IEEEbiography}
\vskip 0pt plus -1fil

\begin{IEEEbiography}[{\includegraphics[width=1in,height=1.25in,clip,keepaspectratio]{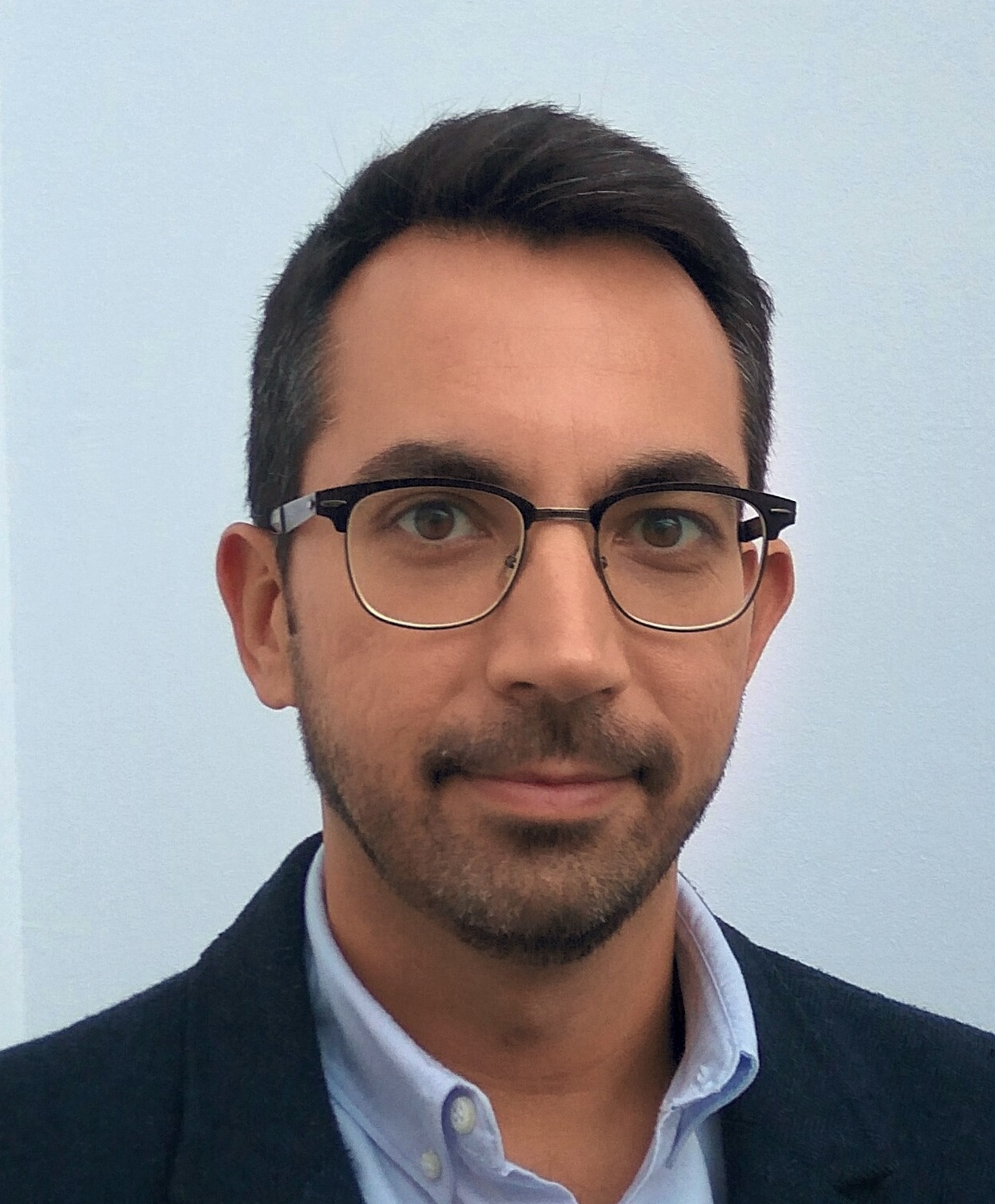}}]{Ruben Vera-Rodriguez}
 received the M.Sc. degree in telecommunications engineering from Universidad de Sevilla, Spain, in 2006, and the Ph.D. degree in electrical and electronic engineering from Swansea University, U.K., in 2010. Since 2010, he has been affiliated with the Biometric Recognition Group, Universidad Autonoma de Madrid, Spain, where he is currently an Associate Professor since 2018. His research interests include signal and image processing, pattern recognition, machine learning, and biometrics. He is the author of more than 130 scientific articles published in international journals and conferences, and 3 patents. He is actively involved in several National and European projects focused on biometrics. He has served as Program Chair for some international conferences such as: IEEE ICCST 2017, CIARP 2018 and ICBEA 2019.
\end{IEEEbiography}

\end{document}